\documentclass[lettersize,journal]{IEEEtran}
\usepackage{amsmath,amsfonts}
\usepackage{algorithmic}
\usepackage{algorithm}
\usepackage{array}
\usepackage[caption=false,font=normalsize,labelfont=sf,textfont=sf]{subfig}
\usepackage{textcomp}
\usepackage{stfloats}
\usepackage{url}
\usepackage{verbatim}
\usepackage{graphicx}
\usepackage{cite}
\usepackage{svg} 
\usepackage{float}
\usepackage{longtable}
\usepackage{tabularx}
\usepackage{multirow}
\usepackage{makecell}

\newcolumntype{C}[1]{>{\centering\arraybackslash}m{#1}}

\newcolumntype{D}[1]{>{\centering\arraybackslash}p{#1}}

\begin{document}

\title{Human Vulnerability Assessment in Cybersecurity: A Systematic Literature Review of Methods, Models, and Instruments}

\author{Dimitra Papatsaroucha, Stavroula Psaroudaki, Eleftheria Vassilaki, \\ Konstantina Pityanou, Michail Alexandros Kourtis, Ilias Politis, \\ Evangelos K. Markakis 
\thanks{Dimitra Papatsaroucha, Stavroula Psaroudaki, Eleftheria Vassilaki, Konstantina Pityanou, and Evangelos K. Markakis are with the Department of Electrical and Computer Engineering, Hellenic Mediterranean University, 71410 Heraklion, Greece \\ 
Michail Alexandros Kourtis is with the National Centre for Scientific Research Demokritos, 15341 Athens, Greece \\ 
Ilias Politis is with the Industrial Systems Institute of the Athena Research Center, 26504 Patras, Greece}
}



\maketitle

\begin{abstract}
In cybersecurity, vulnerability assessment has typically focused on identifying and measuring vulnerabilities within digital assets and technical infrastructures. However, there is growing recognition that this approach alone is inadequate without a structured examination of the human factor, which is becoming more frequently targeted and manipulated by cyber adversaries. Human vulnerabilities extend beyond individual susceptibility to cyber threats, encompassing a wide array of psychological, cognitive, behavioral, social, and contextual factors that can, whether unintentionally or intentionally, jeopardize the security and integrity of systems and data. Despite this recognition, human vulnerability assessment remains fragmented, often addressed from a static rather than a dynamic perspective, and with limited focus on the ways it propagates across individuals and systems; a growing body of literature has explored specific facets of the issue, including one-time assessments of security behavior, user awareness, and, to a degree, intentional insider threats and their detection. This research offers a systematic literature review (SLR) of Human Vulnerability Assessment (HVA) in cybersecurity, including methods, models, and instruments proposed for the conceptual or practical assessment of human vulnerabilities across various dimensions. Following the PRISMA framework, this review gathers relevant studies published from 2017 to 2025, aiming to investigate whether any assessment methods, models, or instruments exist that address the entire spectrum of human vulnerabilities dynamically. The findings highlight gaps and limitations in current proposed solutions and identify areas for further investigation regarding holistic assessment that simultaneously and dynamically considers the entire spectrum of both the unintentional and intentional dimensions of human vulnerability. 
\end{abstract}

\begin{IEEEkeywords}
Human Vulnerability Assessment, Human-Centered Cybersecurity, Human Factors, Insider Threats, Systematic Literature Review.
\end{IEEEkeywords}

\section{Introduction}
\IEEEPARstart{W}{ith} the increasing use of Information and Communication Technologies (ICT) in government, industrial, and organizational contexts as well as in personal environments, cybersecurity has become an essential requirement for modern digital societies. The rapid growth of connected systems, cloud services, smart technologies, and digitally mediated workflows has dramatically expanded the surface of cyber threats. Organizations are increasingly exposed to risks stemming not only from technical vulnerabilities of digital assets and systems but also from the vulnerabilities of their respective human operators, indicating that cybersecurity measures should not be isolated from human behavior, organizational culture, workplace conditions, and broader socio-technical aspects\footnote{https://www.enisa.europa.eu/publications/cybersecurity-culture-guidelines-behavioural-aspects-of-cybersecurity} \footnote{https://www.enisa.europa.eu/publications/cyber-security-culture-in-organisations}. Meanwhile, the Cybersecurity Framework (CSF) 2.0, by the National Institute of Standards and Technology (NIST), highlights people, organizational context, continuous monitoring, and adaptive risk management as key elements of contemporary cybersecurity governance\footnote{https://nvlpubs.nist.gov/nistpubs/CSWP/NIST.CSWP.29.pdf}.

Human error and susceptibility have long been acknowledged as significant factors in cybersecurity risk \cite{Pollini2022, Rahman2021, Amoresano2023}, cotributing to unintentional threats, where users may inadvertently participate in cyber incidents. According to the European Union Agency for Cybersecurity (ENISA), organizations remain vulnerable to cyber threats and data breaches even when technical safeguards and security infrastructures are in place, due to poor cybersecurity awareness, unsafe user practices, and lack of proper security behaviour\footnote{https://www.enisa.europa.eu/publications/raising-awareness-of-cybersecurity}. Meanwhile, cyber attacks are evolving, integrating behavioraly adaptive mechanisms to target human cognition, decision-making, and interaction patterns\footnote{https://www.enisa.europa.eu/news/cybersecurity-threats-fast-forward-2030}. Furthermore, the literature has started to place particular focus on intentional threats \cite{Nurse2014, King2018, Haim2017}, where users may deliberately engage in malicious activities within the system they operate, aiming to exploit it for personal, financial, or idelogical purposes. According to the ENISA Threat Landscape 2025 report, insider threats and AI-assisted malicious activities continue to shape the evolving cyber threat landscape alongside social engineering, such as phishing, that ranks the highest\footnote{https://www.enisa.europa.eu/publications/enisa-threat-landscape-2025}.

Human vulnerability in cybersecurity encompasses a wide spectrum of psychological, cognitive, behavioral, organizational, and contextual human factors \cite{Papatsaroucha2021}. Simultaneously, human vulnerability can be conceptualized in different ways, from static assessment methods that view vulnerability as a relatively fixed individual condition, often evaluated through profiling or psychometric evaluation \cite{Parsons2014, Egelman2015, Cullen2018}, to dynamic and continuous approaches that consider vulnerability as an evolving state affected by behavioral adaptation, contextual conditions, environmental changes, and continuous interaction with digital systems \cite{Miehling2018, Alohali2018, Malek2020}. In parallel, vulnerability may transcend individual isolated conditions and emerge as a broader socio-technical phenomenon involving relational dynamics, influence mechanisms, and propagation effects within and between organizational environments \cite{Perrotin2022}. In this context, Human Vulnerability Assessment (HVA) approaches aim to identify, assess, and model the human conditions that can lead to increased susceptibility to cyber threats.

This study presents a Systematic Literature Review (SLR) of HVA approaches in cybersecurity, using the PRISMA framework \cite{Page2021}. The objective of the review is to systematically investigate whether existing methods, models, and assessment instruments address the broader spectrum of factors affecting human vulnerability while also considering dynamic and continuous assessment perspectives and vulnerability propagation mechanisms. To support this investigation, the study synthesizes the factors and additional variables affecting human vulnerability, as identified across the literature, and proposes a structured taxonomy that is used across the SLR as a classification framework to support deeper analysis of the studies considered eligible for inclusion.

The remainder of this paper is structured as follows: Section \ref{sec:Related Work} presents prior SLRs, surveys, and reviews across the literature that have discussed narrower dimensions of the research objective of this study; Section \ref{sec:Taxonomy} discusses the human vulnerability-related factors identified across the literature and presents a structured taxonomy by grouping them into vulnerability domains and moderator groups; Section \ref{sec:PRISMA} describes in detail the methodology followed in the SLR of this study, in adherence with the PRISMA framework; Section \ref{sec:Methods, Models, Instruments} presents the studies eligible for inclusion, by grouping them into methods, models, and instruments and providing the characteristics of each approach; Section \ref{sec:Results} analyses and synthesises the results of this SLR by classifying the eligible studies according to the proposed taxonomy of Section \ref{sec:Taxonomy} while it also highlights trends, gaps, and limitations of the current landscape; and Section \ref{sec:Conclusion} provides the concluding remarks of this study alongside limitations and future research dimensions.

\section{Related Work}
\label{sec:Related Work}
Prior SLRs, surveys, and reviews across literature have addressed narrower dimensions of the research objective of the present study. Most notably, even though some reviews focus on how cybersecurity-related behavior is assessed, emphasizing the assessment process and its components, they do not synthesize the methods, models, or instruments proposed across the scientific community to assess the full range of human vulnerabilities in cybersecurity contexts \cite{Kannelnning2023}. Other studies review assessment instruments developed for a single construct, such as information security awareness scales, examining their dimensions, rigor, and validity but remaining limited to that specific aspect of human vulnerability \cite{Rohan2023}. At a broader level, some reviews synthesize the human factor in cybersecurity by identifying major vulnerability themes and proposing integrative frameworks, yet without systematically centering the analysis on the methods, models, and instruments used to assess those vulnerabilities \cite{Escobar2025}.

Beyond these studies, additional reviews across the literature explore human factors in cybersecurity by focusing in specific slices of the problem, including particular attack vectors such as cognitive vulnerabilities to phishing \cite{Arvalo2023} and social engineering-related susceptibility patterns, including emotional manipulation, overtrust in authority, and low security awareness \cite{Tsauri2025}, as well as amplification factors such as emotional manipulation and overtrust in AI \cite{Jabir2025}. Other reviews adopt behavioral or theoretical perspectives, examining demographic, psychological, and technical factors influencing security behavior and compliance \cite{Almansoori2023, Alsharida2023, Kuppusamy2022}, or focus on specific constructs such as self-efficacy \cite{Borgert2024}. Several studies further provide interdisciplinary, domain-specific, or context-specific reviews with regard to the involvement or influence of human factors in the security of digital spaces, including insider threats \cite{Abdallah2023, Pathirana2026}, threats imposed by individuals operating within remote work environments \cite{Nizamuddin2025} or in various sectors such as healthcare and agriculture \cite{Nifakos2021, Bissadu2024}, technical detection approaches \cite{Sedraoui2024, Manoharan2024}, and mitigation-oriented perspectives such as awareness training and security culture \cite{Abzakh2023, Rohan2021, Masimba2025}.

Despite the existence of these reviews, the literature still lacks a systematic and comprehensive synthesis of the methods, models, and instruments proposed to assess human vulnerabilities in cybersecurity across their full spectrum. Existing reviews remain inherently scoped to specific dimensions, such as particular behavioral constructs, single vulnerability domains, specific attack vectors, or broader human-factor discussions that do not center on assessment approaches themselves. In contrast, the present study adopts a holistic perspective by systematically identifying and analyzing all methods, models, and instruments proposed in the literature between 2017 and 2025 that aim to assess at least one dimension of human vulnerability in the cybersecurity domain. By consolidating these approaches across psychological, cognitive, behavioral, social, and contextual dimensions, this work provides a holistic overview of the current assessment landscape, enabling a more integrated understanding of how human vulnerabilities are conceptualized and measured.

\section{Human Factors}
\label{sec:Taxonomy}

Across the literature there are partial classifications provided with regard to human factors in cybersecurity. However, to the best of our knowledge, there lacks a unified taxonomy specifically designed for human vulnerability assessment. Some studies propose ontology-based representations of human factors within cybersecurity contexts \cite{Alessandro2015}, while others propose socio-technical assessment approaches integrating human, organizational, and technological aspects that influence vulnerability within the cyber space \cite{Pollini2022}. Additional studies provide conceptual categorizations of human cybersecurity factors and identify recurring categories such as psychological traits, cognitive mechanisms, behavioral practices, and contextual influences \cite{Alessandro2015, Pollini2022, Khadka2025, Rohan2023, Jeong2019}. In addition, the synthesis and categorization of factors proposed in \cite{King2018} has emphasized that malicious cyber behavior emerges from various inter-connected aspects at multiple levels, including individual characteristics, interpersonal dynamics, group affiliations, and broader socio-cultural contexts. 

Nevertheless, none of these studies presents a comprehensive and holistic taxonomy specifically structured for evaluating human vulnerability mechanisms across the full human cybersecurity lifecycle considering both unintentional and intentional threats. Instead, existing classifications often mix intrinsic vulnerabilities with contextual variables, treat behavioral outcomes and underlying psychological mechanisms interchangeably, or focus on specific domains such as intentional insider threats, socio-technical risk analysis, or security awareness.

To address this gap, the present study draws insight from current literature providing such partial classifications as elaborated above, identifies recurring concepts, and merges them in order to develop a Cybersecurity Human Factor Taxonomy designed specifically to support the SLR of HVA methods, models, and instruments. The taxonomy was created by incorporating evidence from two sources following an iterative process to synthesize the concept. The first source was the factor structure presented in \cite{Papatsaroucha2021}, which identified many human susceptibility factors linked to both unintentional and intentional cyber threats. These encompassed demographics, personality traits, cognitive processes, emotions, cybersecurity awareness, training, behavioral patterns, cultural influences, and factors associated with maliciousness \cite{King2018, Henshel2015, Henshel2016, Arend2020, Frauenstein2020, Ovelgnne2017, Whalen2007, Albladi2018}. The second source consisted of the analysis of the candidate taxonomy-related studies that were identified during the preparatory stage of this review \cite{Alessandro2015, Pollini2022, Khadka2025, Jeong2019, Rohan2021}. These studies were examined specifically to determine whether an existing taxonomy could be adopted or extended instead of constructing a new classification. Recurring human factor categories and structural patterns relevant to cybersecurity vulnerability analysis were identified and evidence and insights extracted from these studies have been further enhanced, extended, and combined to construct the taxonomy utilised in the current study. 

\subsection{Taxonomy Design \& Development}

The developed taxonomy has been structured around two overarching classes, namely: i) vulnerability domains as conceptual categories, each of them grouping various human factors that function as vulnerability mechanisms, and the associated sub-factors that portray vectors, dimensions, or manifestations of the corresponding human factor, ii) contextual variables that function as moderators of human vulnerabilities. Taxonomy-related literature often merges these two classes, treating variables such as demographics, training, or cultural environment as vulnerabilities themselves. However, these variables mostly influence how vulnerabilities emerge or manifest rather than constituting vulnerabilities per se. This was particularly evident in relevant literature that has been investigating how such variables influence cognitive- or behavior-related vulnerability levels \cite{Yan2018, Anwar2017, McEvoy2019, Orji2018, Mamonov2018, Wiley2020, Ebner2020, Chen2020}.

The design process adopted commonly utilised taxonomy design principles, identified across the literature \cite{Nickerson2013}, namely: mutual exclusivity, which helps ensure that each identified human factor belongs to only one vulnerability domain and duplication is avoided; collective exhaustiveness, according to which the pool of human factors needs to capture the entire conceptual space of human vulnerability, hence taxonomy related studies as well as studies measuring, assessing, or aiming to identify links between susceptibility and specific human factors were investigated to construct the human factor pool; and conceptual coherence, which dictates that each domain of human factors groups concepts that have a similar functional role in shaping human vulnerability within cybersecurity contexts. These design principles formed the basis upon which four methodological steps were applied to construct the taxonomy.

First, all candidate human factors extracted from the literature were consolidated into a single concept pool. This pool included both unintentional threat concepts, such as risk-taking behavior, lack of awareness, information disclosure practices, and security compliance tendencies, as well as intentional threat concepts, such as malicious personality traits, norms, values, and aggressive interpersonal behavior. Second, all factors of the pool were separated into the two major conceptual classes, by distinguishing between intrinsic vulnerability mechanisms and contextual variables, identifying vulnerability human factors and moderator indicators, respectively. Third, factors and indicators, resepctively, that were similar in concept were combined and restructured to avoid repetition while enhancing clarity. For example, trust-related constructs identified in several studies were combined to form the trust perception human factor, and technology acceptance variables like perceived usefulness, perceived ease of use, and behavioral intention to use systems were combined into technology perception. Fourth, the final list of factors and indicators were grouped into higher-order conceptual vulnerability domains and moderator groups, respectively, based on their functional role in human cyber vulnerability mechanisms.

\subsection{Cybersecurity Human Factor Taxonomy}

The developed Cybersecurity Human Factor Taxonomy follows a causal interaction model \cite{Bandura1999}, commonly discussed in behavioral research, suggesting that psychological factors influence cognitive processes that in turn shape or affect behavioural patterns, while temporal conditions and environmental context may further affect behavior. The taxonomy is structured around i) four main vulnerability domains, reflecting the primary layers through which relevant human factors may be triggered and vulnerabilities may manifest in cybersecurity contexts and ii) four moderator groups, representing contextual conditions that may amplify, mitigate, or shape the effects of the human factors grouped under the four vulnerability domains. Both intentional and unintentional threats are considered in this structure, as human vulnerability extends beyond accidental user error to include factors associated with deliberate misuse of systems and data as well as with the exploitation of vulnerabilities of other users (e.g., through social engineering or insider manipulation) within the same cyber-physical system. This structure highlights, as already recognized across the literature \cite{King2018, Jureviien2021, Ovelgnne2017, Nurse2014, Alessandro2015}, that human vulnerabilities and malicious cyber behaviors emerge from the interaction of factors operating at multiple levels.

Vulnerability domains include Psychological Factors, Cognitive Factors, Behavioral Factors, and Human Performance State Factors Regarding Psychological Factors, these include relatively stable individual characteristics that influence how individuals perceive and respond to cyber threats, such as personality traits, emotions, ethics, attitudes, mental stability, and self-perception. Under this domain, traits such as moral disengagement, low empathy, or malicious personality tendencies may be indicative of potential insider misuse \cite{King2018}. 

Cognitive factors refer to mechanisms employed for information processing, including cognitive processes, biases, awareness, expertise, trust perception, technology perception, and risk appraisal. Within this domain, a key highlight in the taxonomy is the incorporation of emerging cybersecurity challenges associated with human-AI interaction, which is currently lacking from other classifications across the literature, potentially due to the very recent widespread use of Large Language Models (LLMs) in both professional and personal everyday life. Modern cyber environments increasingly involve the use of AI-driven tools, which introduce new forms of cognitive vulnerability such as automation bias, overreliance on AI-generated outputs, and reduced verification of machine-generated information \cite{Romeo2025, Tilbury2026, Teo2025}. These vulnerabilities were incorporated into the taxonomy under the Human-AI Cognitive Interaction factor, as they represent specific manifestations of trust evaluation and cognitive bias in AI-assisted decision-making environments.

On the other hand, behavioral factors describe the action patterns that expose individuals to cyber threats, including security behavior like risk-taking or policy non-compliance, and online behavior, such as information disclosure or browsing habits. In addition, behavioral factors may also capture observable manifestations of intentional threats, such as deliberate policy violations, misuse of access privileges, or malicious online actions. The distinction between cognitive and behavioral factors lies in the fact that cognitive factors capture knowledge, awareness, and perception of safe behavior, whereas behavioral factors reflect the actual actions and practices performed by users in digital environments. Furthermore, human performance state factors represent temporary performance conditions, such as alertness, time pressure, multitasking, and interruptions, which have a situational and dynamic nature. Contrary to stable characteristics, these factors may vary from one day to the next and their influence may change depending on the temporal and operational context of cybersecurity decisions.

The Moderator class includes groups of variables such as demographics, cybersecurity training, prior experience with cyber incidents, and socio-cultural \& environmental context. These further influence how vulnerabilities emerge, i.e., how human factors are triggered or how strongly they manifest, while they also shape the relationship among various human factors. Research indicates that demographics such as age, education, and occupation may have an effect on cybersecurity awareness and risk perception \cite{Wang2021, Parsons2019, Orji2018, Anwar2017, Taib2019}. Furthermore, cybersecurity training may enhance awareness and improve security behavior by enhancing threat detection and response skills \cite{Sumner2022,Cuchta2019, Linkov2019}. Meanwhile, prior experience with cyber incidents may influence how individuals perceive and respond to cybersecurity risks, by either making them overconfident on their defense competency or very reluctant to proceed to risky behaviors \cite{Shakela2019, Chen2020, Jaeger2021}. This includes dimensions such as prior victimization, success or failure of incident detection, exposure only by observation, and previously adopted defense actions.

Socio-cultural and environmental indicators, including norms, values, governance conditions, economic pressure, legal frameworks, and intergroup dynamics, strongly relate to intentional threats, as they may affect the perception of insider threat behavior as unacceptable, tolerated, or justified within a digital environment. Overall, socio-cultural and environmental factors highlight that malicious insiders cannot be fully understood solely through individual characteristics. Instead, intentional malicious cyber behavior often emerges from the interaction between individual traits and the broader social, cultural, and institutional contexts in which digital activities occur \cite{King2018}.

In Table \ref{tab:tabl1}, below, the four Vulnerabilities Domains of the developed taxonomy are provided alongside the human factors they are composed of and several sub-factors that were identified across the literature as investigated dimensions. For each human factor, Table \ref{tab:tabl1} indicates the Threat Relevance as well, that is the potential threat that may emerge from the manifestation of the human factor, i.e., Unintentional Threat (U), Intentional Threat, or both (U/I), as it was identified through the literature. In addition, Table \ref{tab:tabl2}, further below, provides the four Moderator Groups, accompanied by the indicators that comprise them, the human factors that may be affected by their manifestation, as identified through the literature, and the Threat Relevance in a similar manner as in Table \ref{tab:tabl1}. Each Vulnerability Domain and Moderator Group and their respective human factors and indicators are followed by a unique ID, in order to streamline the analysis of the studies included in the SLR and categorize the methods, models, and instruments accordingly in the sections that follow.

Overall, the proposed taxonomy should be understood as a literature-grounded analytical framework developed specifically to support the systematic analysis of HVA-related studies reviewed in the current SLR. It does not attempt to define a psychological measurement instrument itself. Instead, it functions as a structured classification scheme used to map the coverage of existing HVA methods, models, and instruments across the space of human cyber vulnerabilities. Therefore, the purpose of this section is not to provide an exhaustive theoretical treatment of each factor, but to define the classification logic used to map the included studies. By systematically recording which vulnerability domains, human factors, moderators and indicators are addressed by each approach identified in the SLR, the taxonomy enables the review to identify conceptual gaps and determine whether current research adequately addresses the multidimensional nature of human cyber vulnerability, which is the primary objective of this study. 

\clearpage
\begin{table*}[!t]
\centering
\caption{Vulnerability Domains and Human Factors}
\label{tab:tabl1}

\renewcommand{\arraystretch}{1.5} 
\setlength{\tabcolsep}{6pt}       
\begin{tabular}{|C{3cm}|C{3.2cm}|C{6cm}|C{1.5cm}|C{1.8cm}|}
\hline
\textbf{Vulnerability Domain} & \textbf{Human Factor} & \textbf{Subfactors / manifestations} & \textbf{Threat relevance} & \textbf{References} \\
\hline

\multirow{7}{*}{\makecell{Psychological \\ Factors (PsyF)}} 
& Personality traits (PsyF-1) & openness, agreeableness, conscientiousness, neuroticism, extraversion, sensation-seeking, impulsivity; Machiavellianism, psychopathy, narcissism & U / I & \cite{Wang2021, Cullen2018, Uebelacker2014, Curtis2018, Cho2016, Nurse2014, Henshel2015, King2018, Alessandro2015, Henshel2016, Khadka2025, Rohan2021, Jeong2019} \\ \cline{2-5}

& Emotions \& feelings (PsyF-2) & mood, stress, anxiety, fear, anger, curiosity, guilt, regret, tension, happiness, sadness, disgust & U / I & \cite{Henshel2015, Wang2021, Chowdhury2020, Alessandro2015, Mamonov2018, Pollini2022, Khadka2025} \\ \cline{2-5}

& Dispositional characteristics (PsyF-3) & conformity, credulity, friendliness, kindness, humility, courtesy, apathy, indifference, envy, self-control; sympathy, helpfulness, self-love, greed, lust, gluttony & U & \cite{Wang2021, Jeong2019} \\ \cline{2-5}

& Ethics / moral reasoning (PsyF-4) & moral disengagement, moral intensity, personal responsibility & U / I & \cite{Wang2021, AlNuaimi2024, Pollini2022, Gillam2020, Khadka2025} \\ \cline{2-5}

& Mental stability (PsyF-5) & stability, instability & I & \cite{Henshel2015, King2018, Nurse2014, Alessandro2015, Henshel2016} \\ \cline{2-5}

& Self-perception (PsyF-6) & pessimistic, optimistic & I & \cite{Henshel2015, King2018, Nurse2014, Alessandro2015, Henshel2016} \\ \cline{2-5}

& Attitudes toward cyber behaviour (PsyF-7) & positive or negative attitudes toward cybersecurity practices & U / I & \cite{Wang2021, Hadlington2017, Henshel2015, King2018, Nurse2014, Alessandro2015, Henshel2016, Khadka2025, Rohan2021} \\ \hline

\multirow{8}{*}{\makecell{Cognitive \\ Factors \\ (CogF)}} 
& Cognitive processes (CogF-1) & information processing, decision-making / judgment, thinking set & U & \cite{Yan2018, Wang2021, Corbett2015, Alessandro2015, Khadka2025} \\ \cline{2-5}

& Cognitive biases (CogF-2) & reasoning shortcuts, cultural biases & I & \cite{Henshel2015, King2018, Nurse2014, Alessandro2015, Henshel2016, Khadka2025, Jeong2019} \\ \cline{2-5}

& Cybersecurity awareness (CogF-3) & awareness of cyber threats and security practices & U & \cite{Mamonov2018, Shakela2019, Alsharif2021, Pollini2022, Khadka2025, Rohan2021} \\ \cline{2-5}

& Computer expertise (CogF-4) & digital literacy, technical competence & U & \cite{Ovelgnne2017, Hanus2022, Sharma2020, Rohan2021} \\ \cline{2-5}

& Trust perception (CogF-5) & trust in people, automated systems, platforms, and information sharing & U & \cite{Khadka2025, Rohan2021} \\ \cline{2-5}

& Technology perception (CogF-6) & perceived usefulness, perceived ease of use, behavioral intention to use systems & U & \cite{Rohan2021} \\ \cline{2-5}

& Risk \& protection appraisal (CogF-7) & perceived vulnerability, perceived severity, response efficacy, self-efficacy & U & \cite{HasanAsfoor2020, Mwagwabi2018, Liang2010, Scott2015, Alturki2020, DeKimpe2022, Khadka2025, Jeong2019} \\ \cline{2-5}

& Security / privacy perception (CogF-8) & perceived security, privacy concern, perceived control of information sharing & U & \cite{Rohan2021} \\ \cline{2-5}

& Human-AI Cognitive Interaction (CogF-9) & automation bias, AI over-trust, cognitive offloading, prompt-injection susceptibility & U & \cite{Khadka2025} \\ \hline

\multirow{5}{*}{\makecell{Behavioral \\ Factors (BehF)}} 
& Security risk behavior (BehF-1) & active risk-taking, passive risk-taking, policy non-compliance, carelessness, inattentiveness, failure to verify information & U / I & \cite{Weber2002, Arend2020, Kennison2020, Gratian2018, Mwagwabi2018, Gangire2019, Benjamin2017, Pollini2022, Khadka2025, Rohan2021, Kadena2021} \\ \cline{2-5}

& Interpersonal behavior (BehF-2) & hostile aggression, instrumental aggression, relational aggression, indirect aggression & I & \cite{Henshel2015, King2018, Nurse2014, Alessandro2015, Henshel2016} \\ \cline{2-5}

& Online self-disclosure (BehF-3) & disclosure of sensitive and non-sensitive information & U & \cite{Mamonov2018, Zlatolas2019, Khadka2025} \\ \cline{2-5}

& Internet addiction (BehF-4) & normal level, moderate level, severe dependence & U & \cite{Hadlington2017, Rohan2021} \\ \cline{2-5}

& Online / browsing habits (BehF-5) & social media engagement, browsing patterns & U & \cite{Albladi2018, Parker2020, Rohan2021} \\ \hline

\multirow{4}{*}{\makecell{Performance \\ State \\ Factors \\ (PerF)}} 
& Alertness (PerF-1) & cognitive overload, fatigue, tiredness, vigilance level & U / I & \cite{Henshel2015, Wang2021, Chowdhury2020, Alessandro2015, Mamonov2018, Pollini2022, Khadka2025, Kadena2021} \\ \cline{2-5}

& Time pressure (PerF-2) & urgency-driven decision making & U / I & \cite{Pollini2022, Khadka2025} \\ \cline{2-5}

& Multitasking (PerF-3) & simultaneous task handling & U / I & \cite{Pollini2022} \\ \cline{2-5}

& Interruptions (PerF-4) & attention switching, task disruption & U / I & \cite{Pollini2022} \\ \hline

\end{tabular}
\end{table*}
\clearpage

\begin{table*}[!t]
\centering
\caption{Moderators table}
\label{tab:tabl2}

\renewcommand{\arraystretch}{1.5} 
\setlength{\tabcolsep}{6pt}       
\begin{tabular}{|C{3cm}|C{5cm}|C{4cm}|C{1.5cm}|C{1.8cm}|}
\hline
\textbf{Moderator Groups} & \textbf{Indicators / Attributes} & \textbf{Factors potentially affected} & \textbf{Threat Relevance} & \textbf{References} \\
\hline

\multirow{5}{*}{\centering \makecell{Demographics \\ (Dem)}} 
& Age (Dem-1) & \multirow{5}{4cm}{\centering awareness, expertise, risk perception, security behavior} & \multirow{5}{*}{Primarily U} & \\ \cline{2-2}
& Gender (Dem-2) & & & \cite{Wang2021, Parsons2019, Orji2018, Mcgill2018, Anwar2017} \\ \cline{2-2}
& Education level (Dem-3) & & & \cite{Abroshan2021, Ovelgnne2017, Rohan2021, Jeong2019} \\ \cline{2-2}
& Occupation (Dem-4) & & & \\ \hline

\multirow{4}{*}{\centering \makecell{Cybersecurity \\ training  (CTr)}}
& Awareness training (CTr-1) & \multirow{4}{4cm}{\centering awareness, computer expertise, security behavior} & \multirow{4}{*}{U} & \cite{Linkov2019, Wang2021, Cuchta2019} \\ \cline{2-2}
& Formal training (CTr-2) & & & \cite{Parsons2019, Sumner2022} \\ \cline{2-2}
& Certification (CTr-3) & & & \cite{Pollini2022, Khadka2025} \\ \cline{2-2}
& Frequency (CTr-4) & & &  \cite{Rohan2021, Jeong2019}\\ \hline

\multirow{4}{*}{\centering \makecell{Experience with \\ cyber incidents \\ (Cexp)}}
& Prior victimization (Cexp-1) & \multirow{4}{4cm}{\centering awareness, computer expertise, security behavior} & \multirow{4}{*}{U} & \cite{Chen2020} \\ \cline{2-2}
& Prior detection success/failure (Cexp-2) & & & \cite{HasanAsfoor2020} \\ \cline{2-2}
& Observational exposure (Cexp-3) & & & \cite{Jaeger2021} \\ \cline{2-2}
& Defense action adoption (Cexp-4) & & & \cite{Jeong2019} \\ \hline

\multirow{10}{*}{\centering \makecell{Socio-cultural \& \\ environmental \\ context (SCE)}}
& Culture (SCE-1) & \multirow{10}{4cm}{\centering psychological factors, risk perception, disclosure behavior, security behavior} & \multirow{10}{*}{Primarily I} & \cite{Ariu2017}
 \\ \cline{2-2}
& Societal conditions (SCE-2) & & & \cite{Abubaker2016, Wang2021, Wiley2020} \\ \cline{2-2}
& Norms (SCE-3) & & & \cite{Wang2021} \\ \cline{2-2}
& Values (SCE-4) & & & \cite{Frauenstein2020, Akdemir2021} \\ \cline{2-2}
& Economic stability (SCE-5) & & & \\ \cline{2-2}
& Governance environment (SCE-6) & & & \cite{Henshel2015, King2018, Nurse2014} \\ \cline{2-2}
& Media portrayal (SCE-7) & & & \cite{Alessandro2015, Henshel2016, Pollini2022} \\ \cline{2-2}
& Legal status (SCE-8) & & & \cite{Khadka2025, Kadena2021} \\ \cline{2-2}
& Intergroup behavior (SCE-9) & & & \cite{Jeong2019} \\ \hline

\end{tabular}
\end{table*}

Nonetheless, the development of the proposed Cybersecurity Human Factor Taxonomy may be interpreted as a secondary methodological contribution of this study. Beyond its role in supporting the SLR, it may also serve as a reference framework for future research aiming to design more comprehensive human vulnerability assessment methodologies. However, it should be highlighted that this taxonomy should be considered preliminary and additional research and experimentation is required to evaluate its validity before it could be formally considered as a scientifically valid taxonomy.

\section{PRISMA Methodology}
\label{sec:PRISMA}
This paper systematically analyzes a complete set of publications from 2017 to 2025 that were retrieved from various scientific libraries and publication search engines. The core objective of the SLR in this work is to investigate recent and relevant literature to evaluate if existing HVA methods, models, or instruments, either implemented or proposed on a conceptual basis, holistically and dynamically assess human vulnerabilities that may impose unintentional or intentional threats in digital environments. The SLR is guided by the following three primary Research Questions (RQs):

\begin{itemize}
\item{RQ1: What methods have been suggested for assessing human vulnerability in cybersecurity?}
\item{RQ2: What models have been currently proposed or implemented for identifying and evaluating or measuring human vulnerability in cybersecurity?}
\item{RQ3: What human vulnerability assessment instruments, such as questionnaires and scales, have been developed or designed within the realm of cybersecurity?}
\end{itemize}

Therefore, the contributions of this SLR include: i) performing a thorough literature review and identifying proposed approaches towards human vulnerability assessment within a cybersecurity context; ii) synthesizing and presenting the results of the literature review on current methods, models, and instruments for human vulnerability assessment; iii) conducting an analysis of the proposed approaches for human vulnerability assessment, and categorizing these according to the Cybersecurity Human Factor Taxonomy outlined in Section \ref{sec:Taxonomy}; iv) documenting any gaps or limitations observed through the review, analysis, and synthesis of the studies eligible for inclusion.
This section describes the roadmap employed for conducting the SLR. The method is founded on the principles and guidelines of the PRISMA framework \cite{Page2021}.

\subsection{Protocol and Eligibility Criteria}
Publications were deemed eligible for inclusion in this SLR if they were published between 2017 and 2025, through a peer-review procedure, and had the full-text version of the publication available online. Various fields of study, such as sociology, psychology, computer science, and criminology, have shown a keen interest in human vulnerabilities pertaining to deception and fraudulent tactics, commonly employed in cyber attacks. Consequently, for this study, suitable studies were those that concentrated on the formulation of HVA frameworks within the domain of cybersecurity. Additionally, emerging directions, such as adaptive user modeling, AI-driven profiling, personalized cybersecurity education, and LLM-based behavioral modeling, were acknowledged as adjacent spaces to HVA but beyond the practical scope of this review that focused on studies that explicitly operationalize fundamental concepts of HVA.

\subsection{Information Sources \& Search String}
A search query was designed and employed to perform a cross-search across four scientific digital libraries, namely: IEEE Xplore Digital Library\footnote{ieeexplore.ieee.org/Xplore/home.jsp}, ACM Digital Library\footnote{ dl.acm.org/ }, ScienceDirect Digital Library\footnote{sciencedirect.com/}, and SpringerLink Digital Library\footnote{link.springer.com/ }. In addition, the search process was complemented by applying the search query to Scopus\footnote{scopus.com/} and Google Scholar\footnote{scholar.google.com/} scientific search engines in order to identify any relevant publications that were not included in the aforementioned digital libraries. The query underwent minor adjustments, before applying to each digital library and search engine, to ensure that it would be inline with their query syntax guidelines and yield all relevant publications.

\textit{("cybersecurity" OR "information security") AND ("human vulnerabilities" OR "human factor vulnerabilities" OR "human risk factor" OR "user awareness") AND ("assessment" OR "framework" OR "model" OR "tool" OR "scale" OR "technique" OR "strategy" OR "approach" OR "method" OR "UEBA" OR "user behavior analytics") NOT (“training” OR “case study”)}
 
\subsection{Inclusion and Exclusion Criteria}
Inclusion and exclusion criteria were established for the screening process to ascertain that the selected articles are pertinent to the RQs and the main objective of the SLR. Publications were considered eligible for inclusion if they met the study's protocol and eligibility criteria and addressed at least one of the specified RQs. 
Consequently, research studies were rejected based on the following exclusion criteria: i) Studies failing to meet the eligibility criteria of the SLR, specifically those published outside the 2017-2025 timeframe, lacking full-text online availability, not peer-reviewed, or not directly pertinent to the cybersecurity domain; ii) studies irrelevant to the RQs of the SLR; iii) studies not authored in the English language; iv) duplicate and repetitive studies; v) studies not directly addressing human vulnerabilities or focusing exclusively on technological vulnerabilities, or theorizing the concept of human vulnerability assessment in cybersecurity without proposing a specific method, model, or instrument for evaluating the human factor within the cybersecurity context.

\subsection{Article Selection \& Search Results}
\begin{figure}[H]
    \centering
    \includegraphics[width=1.0\linewidth]{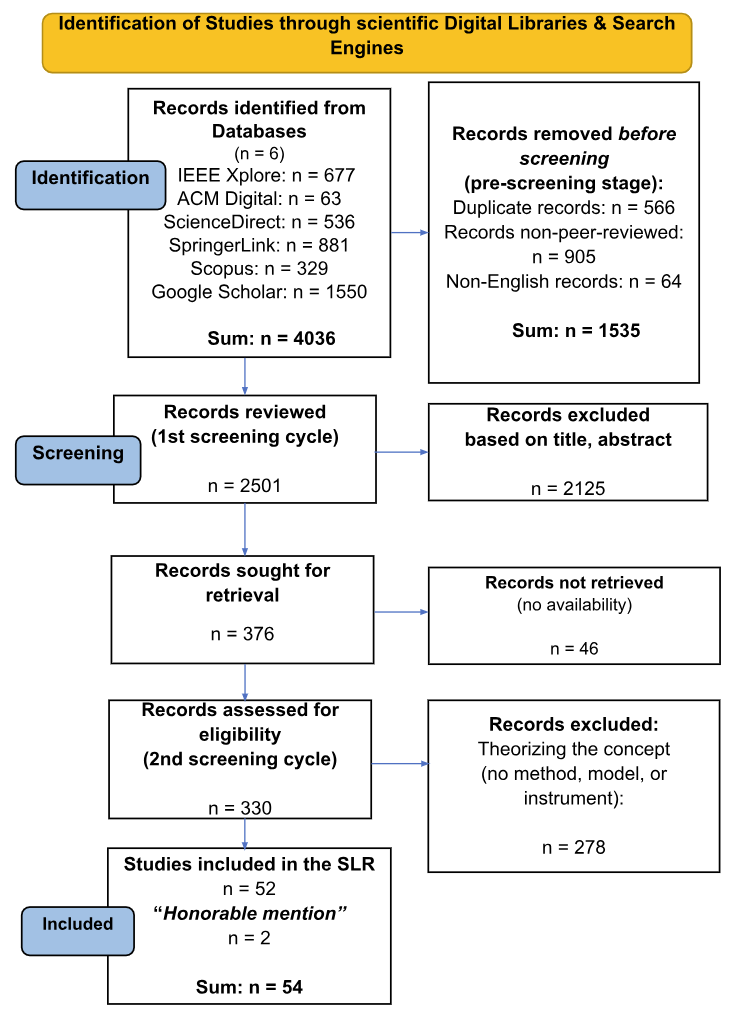}
    \caption{Identification of Studies following the PRISMA framework}
    \label{fig:fig0}
\end{figure}

The search yielded 4,036 studies, which were reviewed by three independent researchers to eliminate bias, in accordance with the procedure depicted in Figure \ref{fig:fig0}. At the pre-screening stage, 1,535 records were excluded according to the exclusion criteria of duplicate records, publication period, peer-review status, and manuscript language. The studies stemming from the pre-screening stage underwent two screening cycles, during which they were categorized into three groups according to the inclusion and exclusion criteria: included, excluded, and maybe. The papers categorized as "maybe" underwent additional scrutiny and assessment by the researchers to decide upon their inclusion or exclusion.

During the 1st screening cycle, 2,501 articles were evaluated, and 2,125 were excluded based on title and abstract according to the established inclusion and exclusion criteria, leading to 376 studies sought for retrieval. Out of these, 46 could not be accessed owing to lack of online availability, such as \cite{Luo2025,Girma2025,Omotoye2026}, which were considered as potentially relevant works in the field of human vulnerability assessment in cybersecurity based on their abstracts, but could not be accessed due to subscription limitations.

During the 2nd screening cycle, 330 retrieved studies were evaluated based on the established protocol and eligibility criteria, leading to the exclusion of 278 studies that focused on theorizing human vulnerabilities instead of proposing a specific, either implemented or conceptualized, method, model, or instrument for assessing human vulnerabilities in a cybersecurity context. 
At the end of the entire screening process, 52 articles were selected for inclusion in this SLR, complemented by 2 additional studies, published in 2015 \cite{Egelman2015} and 2014 \cite{Parsons2014}, which were identified through a backward snowballing process and acknowledged as "honorable mentions" due to their esteemed recognition within the scientific community and their substantial citation records.

\begin{figure}[H]
    \centering
    \includegraphics[width=0.9\linewidth]{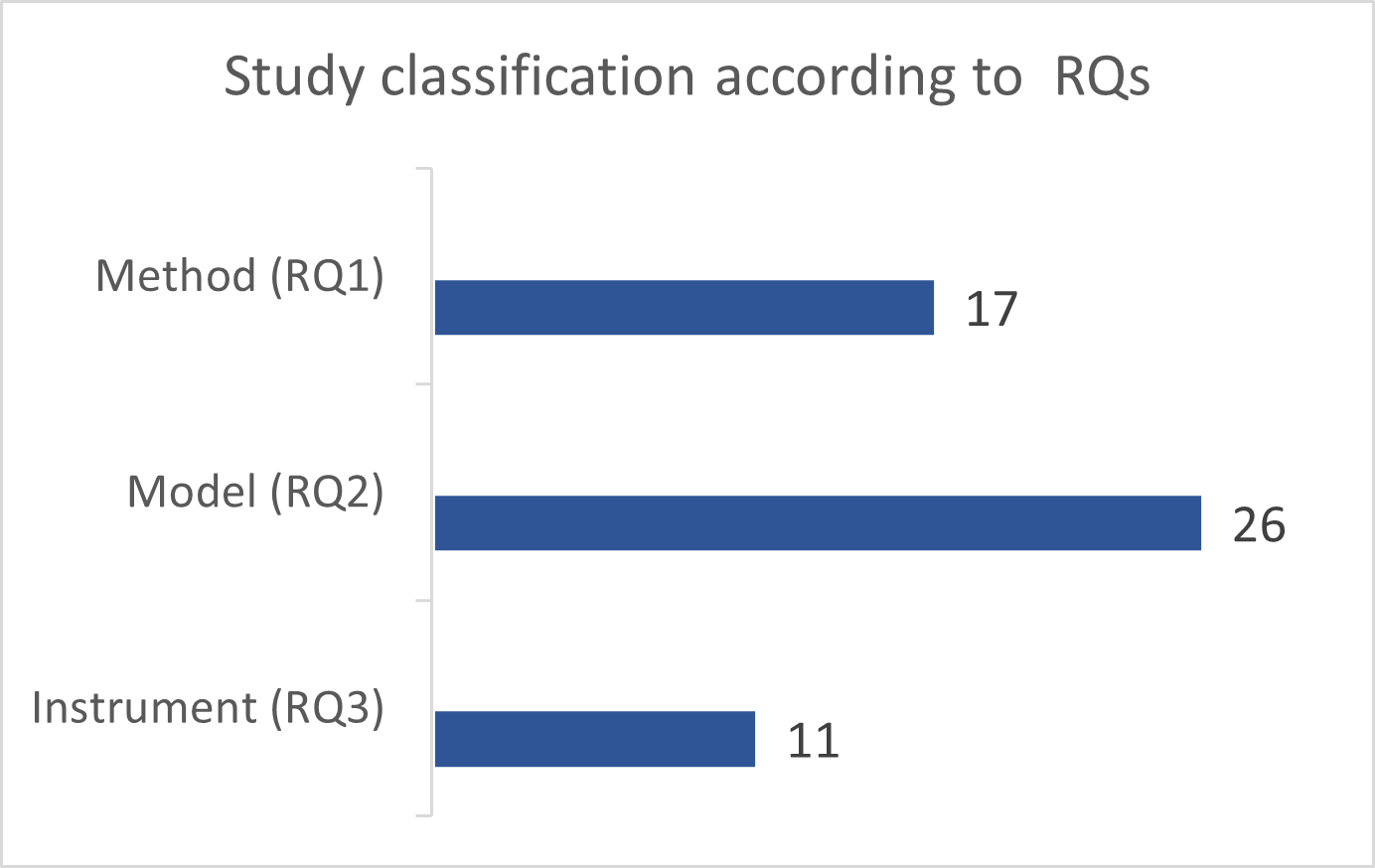}
    \caption{Study classification according to RQs}
    \label{fig:fig1}
\end{figure}

The classification of the selected studies according to the RQs to which they relate to is illustrated in Figure \ref{fig:fig1}. The research process revealed that most studies, 26 in particular, focus on designing and implementing computational models to measure dimensions of human vulnerability in cybersecurity. This is followed by 17 studies that propose and apply a method or framework. Finally, 11 studies present instruments, such as quastionnaires and scales, to assess specific human vulnerability factors.

\begin{figure}[H]
    \centering
    \includegraphics[width=0.9\linewidth]{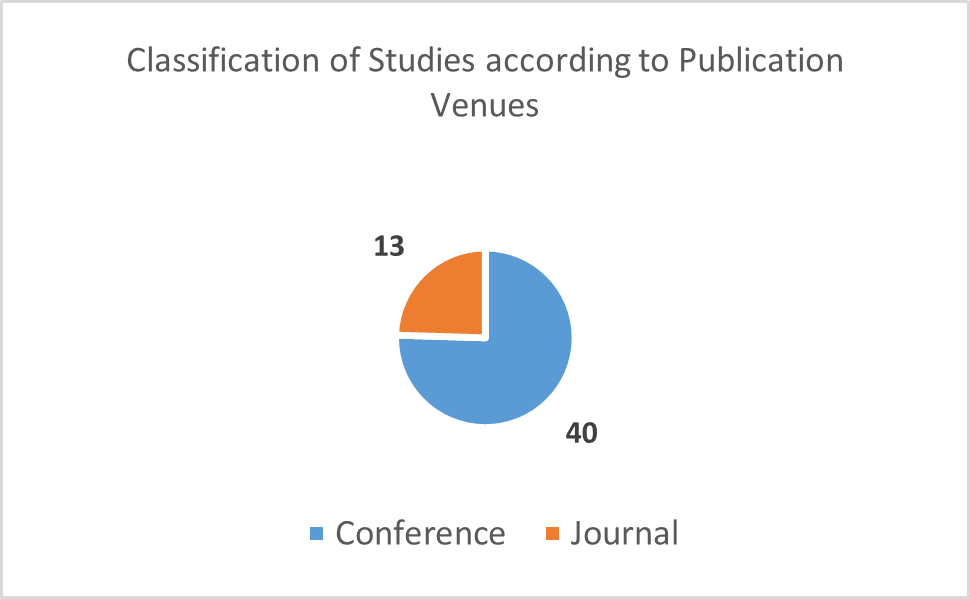}
    \caption{Classification of Studies according to Publication Venues}
    \label{fig:fig2}
\end{figure}

Figure \ref{fig:fig2} provides information about the publication venues of the selected studies. As illustrated, the majority of studies, 40 in prticular, originated from well-established conferences, while the rest 13 of them, originated from high-impact and peer-reviewed journals.

\begin{figure}[H]
    \centering
    \includegraphics[width=0.9\linewidth]{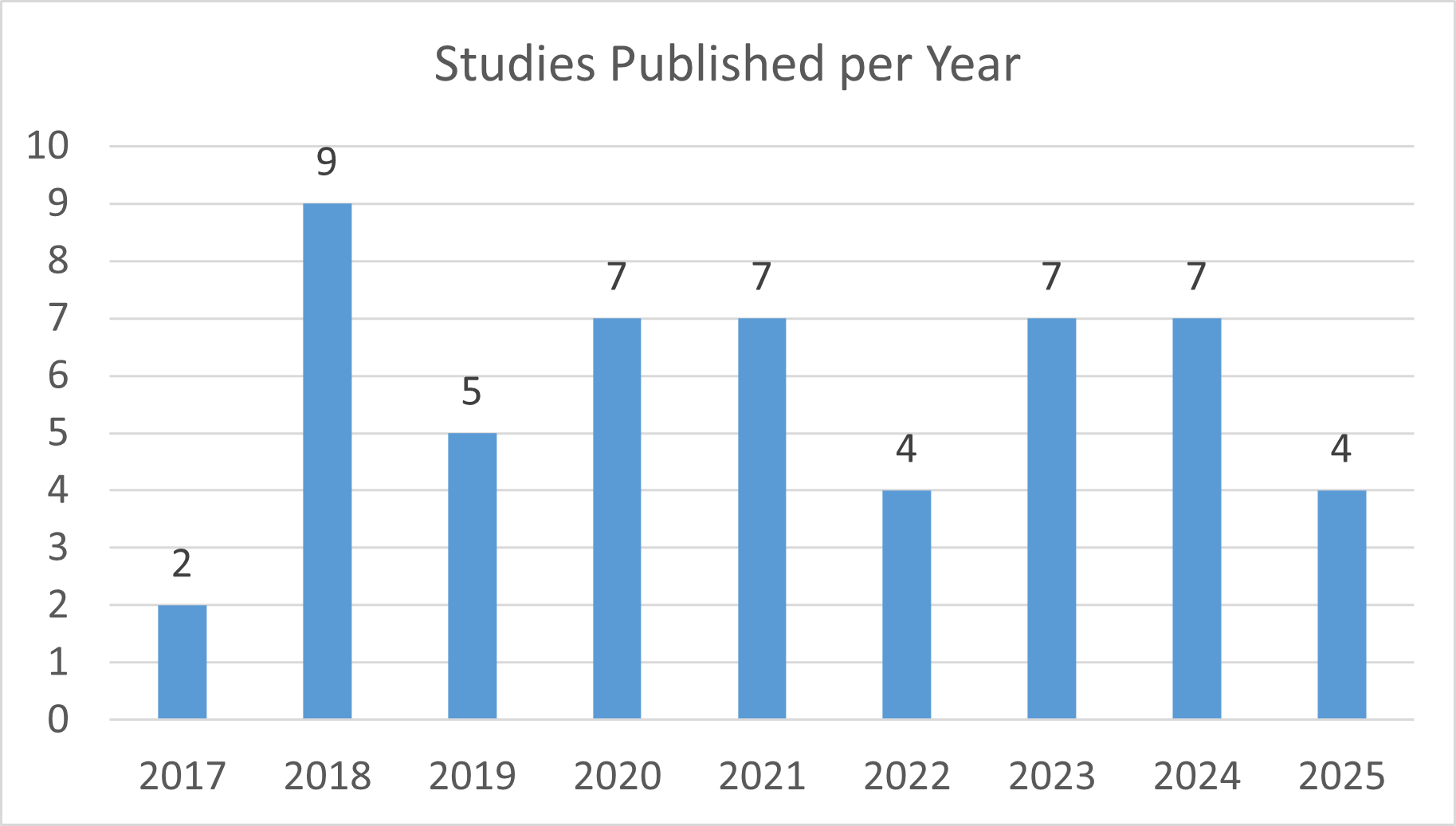}
    \caption{Studies Published per Year}
    \label{fig:fig3}
\end{figure}

Regarding the publication period of the included studies, the distribution of studies generally increases over time as depicted in Figure \ref{fig:fig3}. As the SLR focused on the period 2017-2025, the studies of 2014 \cite{Parsons2014} and 2015 \cite{Egelman2015}, acknowledged and included in the SLR as “honorable mentions”, were excluded from this diagram as they do not reflect the full publication activity of those years.

Focusing on the main study period, the research activity begins with a moderate number of 2 studies in 2017, followed by a sharp rise in 2018 with 9 studies, which also represents the year with the most published studies. From 2019 and onwards, the number of studies is somewhat stabilized at a moderate level, where 5 studies were published in 2019, and 7 studies were published in both 2020 and 2021. A small and temporal decline is observed in 2022, with 4 studies were published in that year, followed again by an increase in both 2023 and 2024 where 7 studies were published. The most recent year, 2025, shows a decrease similar to 2022, with 4 studies in that year. Overall, the figure suggests that the topic of human vulnerability assessment research has gained significant attention after 2018 and remains consistent, despite minor fluctuations.

\section{How Human Vulnerability is Assessed in Cybersecurity}
\label{sec:Methods, Models, Instruments}
\subsection{Methods}
\subsubsection{Conceptual and Theoretical Methods}
Conceptual and theoretical methods have mostly focused on defining how human vulnerability may be understood before proposing practical assessment mechanisms, providing important explanatory foundations and identifying important vulnerability variables. For unintentional threats, emphasis is placed on awareness degradation, cognitive overload, and policy noncompliance. As proposed by \cite{Li2024}, linking security communication fatigue to information security policy noncompliance through the Protection Motivation Theory can provide insights on risk perception and self-efficacy under cognitive exhaustion. Furthermore, the BYOD-SAM framework proposed by \cite{Shihepo2023} relies on technical controls and IT monitoring to assess and improve compliance of enterprise mobile-device users. For intentional threats, focus has been primarily on insider-threat dynamics, social influence, and psychological predispositions. Building on the Motivation, Opportunity, Capability (MOC) model, \cite{Chaipa2023} proposed the MOCR theory, incorporating rationalization and influence / societal conditions from fraud theory to explain how an insider's risk profile may evolve into malicious behavior. In addition, cognitive psychology dimensions have been utilized by \cite{Kim2023} to classify cybercognitive attacks through factors such as cognitive bias, mental stability, and personality traits, to explain how human vulnerabilities may be exploited to bypass technical defenses.

Such approaches emphasize the psychological, cognitive, and organizational conditions influencing vulnerable or malicious cyber behavior; however, most remain pre-empirical and lack standardized assessment metrics, practical validation, and mechanisms for continuous or adaptive assessment.

\subsubsection{Behavioral Observation and Analytical Methods}
Behavioral observation and analytical methods conceptualize vulnerability primarily through observable actions, behavioral traces, and measurable indicators. Instead of relying on perceived attitudes or self-reported awareness, these approaches have attempted to assess vulnerability objectively through monitoring systems, scoring mechanisms, and quantitative analysis. For instance, the Pattern Based Intrusion Detection (PIDE) method continuously monitors user activity and compares it against expert-defined authorized and unauthorized behavioral signatures in order to identify suspicious insider actions \cite{Malek2020}. From the analytical perspective, the inclusive Social Cyber Vulnerability (iSCV) metric was used by \cite{Mitra2024} to evaluate the susceptibility of underrepresented populations to social cyberattacks by analyzing demographics, computer literacy, and previous attack exposure. Regarding combining methods, the Digital-PASS simulation and the SocialScore tool were combined by \cite{McHatton2023} to quantify risky social media exposure and careless online self-disclosure through gamified privacy education mechanisms.

Such approaches provide scalable and objective assessment mechanisms that are often compatible with continuous monitoring environments; however, the fact that they often rely on fixed behavioral signatures may render them difficult to adapt to previously unseen patterns. At the same time, analytical scoring systems may lack the psychological depth necessary to explain why users exhibit certain vulnerabilities.

\subsubsection{Self-Reported and Qualitative Methods}
Instead of relying on behavioral monitoring, self reported and qualitative approaches attempt to capture vulnerability through direct user input or contextual expert analysis. Self-reported methods conceptualize vulnerability primarily through perceived experiences, attitudes, and beliefs while qualitative and expert-based methods are mainly used to explore complex organizational and psychological dimensions of vulnerability, which can be particularly useful for capturing psychological states, perceptions, and contextual experiences. 

With regard to unintentional threats, self-reported methods have been used to assess knowledge, attitude, and behavior (KAB) dimensions of the cybersecurity maturity of users in \cite{Barath2025} while questionnaires and self-diagnosis tools have been combined to identify high-risk behaviors before administering targeted awareness interventions in \cite{Antunes2021}. A four-component model of cyber fatigue was proposed and evaluated through qualitative case studies and interviews, assessing and categorizing fatigue into advice / action-related and attitudinal / cognitive dimensions \cite{Reeves2021}. In addition, graph-based scenario modeling and cognitive maps have been utilized by \cite{Feyzov2023} to forecast organizational susceptibility to Business Email Compromise by cross-referencing personality traits against emotional vulnerabilities across departments.

Such methods remain limited by subjectivity, self-report bias, and their reliance on perceived rather than actual security behavior under realistic operational conditions.

\subsubsection{Experimental and Simulation-Based Methods}
Experimental and simulation-based methods conceptualize vulnerability primarily through user reactions under simulated operational conditions rather than through self-perception or theoretical profiling. They target vulnerability assesmment mostly through exposing users to controlled cyberattack scenarios and capturing realistic behavioral responses. For instance, a Bayesian Multi-armed Bandit testing strategy was proposed in \cite{Miehling2018}, according to which system administrators distribute simulated malicious messages and binary user responses, such as clicking or ignoring the content, are recorded to dynamically identify users who are most susceptible to social engineering attacks.

While such methods exhibit the ability to observe realistic user behavior in controlled attack environments, they often simplify vulnerability into limited behavioral outcomes and may overlook broader psychological, organizational, or contextual factors influencing user decisions.

\subsubsection{Hybrid Methods}
Hybrid methods have attempted to overcome the limitations of single-source assessment by combining multiple data collection and evaluation techniques. These approaches conceptualize vulnerability as a multidimensional phenomenon that cannot be captured adequately through isolated psychometric, behavioral, or technical measurements. For unintentional threats, hybrid approaches frequently have combined psychometric profiling with behavioral experimentation or continuous monitoring, such as integrating MBTI personality assessments with simulated phishing experiments, to demonstrate how different personality types respond differently to social engineering tactics \cite{Cullen2018}. Similarly, questionnaires have been combined with AI-driven image and text analysis to construct a "Digital Human" ontology capable of generating personalized vulnerability scores from inferred human traits \cite{Jureviien2021}. Vulnerability has also been assessed through continuous monitoring of user activity combined with demographic characteristics, IT expertise, and personality-related information \cite{Alohali2018}. For intentional threats, hybrid methods have integrated behavioral monitoring with advanced analytics to infer psychological characteristics from user activity. Natural Language Processing (NLP) has been applied to employee-generated text in order to infer Big Five personality traits, which are subsequently used by machine learning algorithms to classify insider-threat profiles such as disgruntled employees or inadvertent insiders \cite{Eftimie2020}.

 The trade-off for these approaches lies in complexity, dependence on continuous data collection, and integration of AI-driven analytics, which introduce important challenges regarding scalability, privacy, explainability, and practical deployment.

\subsection{Models}
\subsubsection{Machine learning-based Models}
Machine-learning techniques have been employed across studies to detect intentional threats by identifying anomalies in user’s behavior, typically assessing individuals indirectly by analyzing their browsing habits and system interactions. In most cases, a baseline of normal behavior is established and then mistakes and deviations are detected. An LSTM-CNN framework utilizing the Long Short-Term Memory (LSTM) and Convolutional Neural Networks (CNN) is used to extract features and patterns from system usage in \cite{Shi2018}, while another framework modeled an LSTM-based autoencoder capturing session activities in \cite{Balaram2020}. For improving detection outcomes in unbalanced datasets, a multi-modal user behavioral analytics (UBA) model is suggested in \cite{Kim2019}, incorporating four different detection algorithms. An alternative approach adopts a more network-centric perspective, applying Deep Neural Networks to network traffic parameters and creating behavioral profiles based on browsing activity and device usage \cite{Pan2021}. In addition, traditional machine learning techniques have been used, including data-layered systems combining statistical methods and classifiers for abnormal behavior detection \cite{Zhu2021}, as well as decision trees and random forests for identifying malicious activities through system-level behavior analysis \cite{Dixit2024}.

Such models have been reported as effective, but still showcase several important limitations, such as limited practical applicability due to dependence on synthetic datasets and a focus on detection accuracy at the expense of exploring psychological, behavioral, and social factors affecting the user’s vulnerability in cybersecurity contexts. As a result, such solutions seem to provide fragmented information about the root causes of insecure behavior, while their evaluation is often limited and underreported, with minimal to no insight into metrics and dataset characteristics, thus restricting precision and comparative analysis.

\subsubsection{Analytical / Quantitative Models}
Unlike machine learning approaches that have focused primarily on anomaly detection, analytical and quantitative models have targeted to explain and predict susceptibility through formal modeling, mathematical analysis, and statistical techniques. Several approaches have focused on phishing susceptibility and cognitive weaknesses associated with unintentional threats. Phishing success probability is estimated through the analysis of behavioral and emotional characteristics in \cite{Duman2023} by focusing particularly on the "hooking factor", while a multi-layered statistical regression model, proposed in \cite{Duman2025}, combines psychological state, digital literacy, and cognitive biases to generate cumulative vulnerability ratings related to phishing susceptibility. Other approaches have translated risky user behavior into quantitative risk scores through telemetry data and weighted behavioral indicators \cite{Wijesinghe2024}. More psychometric-oriented models, like the Implicit Association Test-based model proposed in \cite{Mocerino2024}, measures unconscious cognitive associations related to phishing vulnerability through reaction-time scoring and urgency or persuasion cues, whereas fuzzy-set Qualitative Comparative Analysis has been used in \cite{Adnan2025} to identify combinations of factors such as security education and awareness associated with vulnerability in e-commerce settings.

Although these approaches provide more human-centric and measurable representations of vulnerability, many rely on small controlled samples, assume relatively static relationships between factors, and remain limited in their ability to capture the dynamic and context-dependent nature of human behavior in operational cybersecurity environments.

\subsubsection{Behavioral Observation / Monitoring Models}
In contrast to both machine learning-based and analytical models, some studies have focused on the direct observation and monitoring of user behavior to detect anomalies. By tracking user activities and system interactions over time, such models evaluate human vulnerability in real-world environments. These approaches collect and analyze behavioral data to identify deviations connected to security threats, instead of directly assessing susceptibility or relying purely on anomaly detection. With regard to intentional threats, the use of UBA in \cite{Haim2017} and User and Entity Behavior Analytics (UEBA) in \cite{Rengarajan2021} operate by establishing baselines for normal user behavior and flagging deviations by tracking user activities in organizational networks. User actions, such as access patterns and logins, form behavioral profiles that are displayed in dashboards and help identify insider threats focused on intentional risks like malicious or rule-breaking behaviors, assisting security analysts to evaluate responses and increase the effectiveness of the systems in operational settings \cite{Rengarajan2021}. In other approaches, the aggregation of user activity over a defined time frame has been utilized in order to minimize the false positives in anomaly identification \cite{Jiang2018}. Systems incorporating behavioral biometrics (BB), device-level tracking, and open-source intelligence (OSINT) have enabled behavior analysis to assess vulnerabilities of malicious intents \cite{Astakhova2019}.  Significantly, other models have focused on personal and organizational dimensions of vulnerability, promoting compliance by prioritizing policy awareness, training, and security proficiency \cite{Alhosani2019}. 

The behavioral monitoring models that have been proposed across the literature showcase practicality and real-time capabilities but they also have key limitations. Many rely on case studies or small experiments with unclear metrics, lacking rigorous quantitative evaluation, while they depend on strict guidelines making them difficult to adjust to evolving threats and behaviors. While implementation remains limited to particular situations, it is worth noting that there is a focus on log-derived behaviors, which provides vague insights on the user’s psychological and cognitive  motivations.

\subsubsection{Experimental / Simulation-based Models}
A different set of approaches uses supervised experimental and simulation-based models to assess human vulnerability under specific circumstances. These models have targeted to directly evaluate the relationships between security outcomes and human factors by actively exposing and monitoring users to simulated scenarios such as phishing or social engineering attacks. A key strategy in those approaches is the simulation of social engineering attacks imitating actual danger scenarios. More precisely, the Social Driven Vulnerability Assessment (SDVA) framework, proposed within the DOGANA project\footnote{https://cordis.europa.eu/project/id/653618}, follows an organized methodology that includes data collection, attack planning, simulated attack execution, and post-attack analysis. This approach enables a comprehensive assessment of user vulnerability to social engineering attacks by implementing persona modeling, communication tactics, and contextual delivery factors. Instead of focusing on intentional maliciousness, it exposes responses to situational cues, manipulation, and persuasion, recognizing cognitive and behavioral vulnerabilities \cite{PachecoB}. Similarly, the Spear Phishing Exposure Level (SPEL) framework determines a Threat Exposure Level, assessing user awareness and experience by exposing users to simulated phishing scenarios. Using Protection Motivation Theory, these methods assess self-efficacy and perceived threats through users' failure to recognize or respond to scams \cite{Shakela2019}. A more recent approach presented an RPA-powered phishing campaign simulation platform, capturing measurable behavioral outcomes to phishing attempts \cite{Papatsaroucha2025}.

Nonetheless, most of the models under this category rely on simulated environments and may not capture the complexity of real-world attacks. Additionally, some of these models rely on self-reported or survey-based metrics, misleading real user activity. Some frameworks remain under development or lack quantitative assessment of their efficacy, while validation is partially restricted to certain industries. 

\subsubsection{Expert-based / Qualitative Models}
A different viewpoint is offered by the studies that are using expert judgment and domain expertise to develop structured models for the assessment of human vulnerabilities. These approaches rely heavily on the use of expert’s knowledge to identify and evaluate human-related risk factors. In this context, a model combining the Human Factor Analysis and Classification System  and Fuzzy Fault Tree Analysis \cite{Lin2023} quantifies human error probability in smart grid substations. For human error calculation, the model defines and weighs the contributing factors in the faults tree structure based on expert opinion, calculating the probability of human error based on the expert-based evaluations of cognitive processes, organizational settings, and operational factors. Even though this category includes a single study across the literature, it emphasizes how critical it is to include expert knowledge in structured vulnerability assessment models.

\subsubsection{Self-reported (survey-based) Models}
Similarly to the assessment instruments, self-reported models measure human vulnerability by combining user’s opinions, attitudes, and self-assessed cybersecurity skills. From this perspective, the Human-Factored Cyber Security Capability Evaluation approach aims to identify the weak-link users by measuring workforce cybersecurity knowledge and abilities \cite{Ani2019}. The framework consists of five levels: defining skills and baselines, collecting data via questionnaires, computing capabilities scores, visualizing results and identifying the weakest link. Individual cybersecurity skills are measured by analyzing responses and taking into account factors such as awareness, abilities, behavior, training and demographics. 

Although this method is effective for drawing attention to perceived weaknesses, it is limited in its ability to be generalized due to its reliance on self-reported data, assumptions for the correlation of knowledge, and actual capabilities and validation on a small industrial sample.

\subsubsection{Hybrid Models}
In an effort to provide a more comprehensive understanding and evaluation of human vulnerability, hybrid models have incorporated various approaches, bridging the gap between behavioral monitoring and subjective self-reports as well as between experimental / simulation-based models and real-world data. For instance, the PoinTER framework used GDPR-compliant human pentesting in SMEs to assess vulnerability to phishing deception and cyber hygiene through simulated attacks combined with expert feedback from interviews \cite{Archibald2018}. Although scalability and consistency remain a limitation, this approach enhances interpretability by enabling both behavioral observation and contextual understanding via the interviews. Combining behavioral tracking, survey responses, and contextual elements including location, time, and past browsing activity, \cite{Solomon2022} synthesized a context-based Information Security Awareness (ISA) model that dynamically evaluates user risk in cybersecurity contexts. Even though integration complexity and data alignment create significant obstacles, this model reduces self-report bias by maintaining scalability through the combination of behavioral data with self-reported awareness metrics for comparing perceived activity. Moving one step further, the smartphone based ISA framework in \cite{Bitton2020} combines questionnaires, network traffic monitoring, and simulated cybersecurity challenges to identify vulnerable users and evaluate their responses to attack scenarios. Although a complex experimental design and data collection are needed, this approach combines various assessment dimensions for deepening human assessment. In addition, the HoS-ML framework, modeling human vulnerabilities in socio-technical systems, incorporates behavioral data, psychological / relational factors, and machine learning to simulate risk propagation across organizational roles \cite{Perrotin2020, Belloir2022}. This approach, which reflects a shift toward more intelligent and adaptive models, merges behavioral analysis with machine learning aiming to identify vulnerability patterns.

\subsection{Instruments}
\subsubsection{Self-reported survey-based instruments}
Most identified instruments rely on self-reported survey data and aim to measure cybersecurity awareness, compliance with cybersecurity policies, susceptibility to persuasion, and both intentional and unintentional behaviors that may lead to cybersecurity compromise, such as \cite{Alissa2018}, \cite{Gangire2021}. The common theme across all is their structured questionnaires, usually with Likert-scale items, that aim to measure human vulnerability from the reported knowledge, attitudes, intentions, traits, or practices. Some instruments address broad organizational cybersecurity behavior in multiple domains, such as the HAIS-Q questionnaire \cite{Parsons2014}, which measures knowledge, attitude, and behavior in internet use, email use, password management, incident reporting, information handling, and mobile computing. Other approaches are more specific, such as SeBIS \cite{Egelman2015}, which focuses on password generation, updating, device security, and proactive awareness. Additional instruments adopt more focused vulnerability measurements, such as susceptibility to persuasion included in StP-II \cite{Modic2018}, motivation to comply with security policies through competence, autonomy and relatedness, and cybersecurity failures through disclosure and intrusion vulnerability included in CSEC \cite{Schoenherr2021}, or privacy exposure in the physical device context included in ODPS \cite{Farzand2024}.

Those instruments measure psychometrics and rely on item generation, expert review, and statistical validation through exploratory or confirmatory factor analysis. Their main advantage is scalability since they are easy to be deployed across large samples and enable a standardized comparison between different user groups and settings. However, because they depend on self-report, they assess reported vulnerability rather than actual vulnerability. Responders are frequently asked to indicate what they believe they do, what they intend to do, or what they believe secure behavior is. This inevitably creates the limitation that the responses may indicate perceived norms, desirable behavior, or idealized self-presentation rather than the actual behavior of the user in real vulnerability conditions. Therefore, these instruments can be useful to measure perceived awareness, compliance perspective, and personality traits, but may be less successful at capturing cybersecurity decisions made under pressure, distraction, uncertainty, or deception.

\subsubsection{Gamified / Interactive instruments}
To address the limitations of self-reported instruments, gamified and interactive instruments have utilized scenario-based settings in which users are asked to respond to simulated scenarios rather than hypothetical questions. These approaches aim to evaluate more realistic behavior by integrating cybersecurity decisions within interactive environments. Following the research regarding human behavior toward cybersecurity policies instrument \cite{Alissa2018}, the authors transformed the defined questionnaire into a gamified approach with intelligent, scenario-driven systems, allowing the users to engage with simulated cybersecurity contexts developed through web technologies \cite{Alissa2021}. Their methodology contains additional contextual factors that may be difficult to be measured by surveys, such as workload, interruptions, prior experience, and defense action adoption. 

While this approach reduces the response bias and can better simulate actual user behavior, it is still limited by predefined scenarios and needs validation against real-life environments.

\subsubsection{Hybrid Instruments}
To bridge the gap between self-reported responses and actual behavior, hybrid instruments integrate psychometric data with behavioral evaluation. The BCISQ instrument \cite{Velki2019} combines self-assessment with simulated risky behavior to capture both cognitive and behavioral metrics. Furthermore, an additional approach proposed in \cite{Hussain2024} combines surveys with simulated attack scenarios, including phishing and ethical hacking experiments to indicate that psychological drivers like curiosity and impulsivity can lead users to exhibit risky cyber behavior. These studies highlight the commonly seen disparities between reported awareness and actual behavior, using factors such as awareness and trust perception, which can significantly influence user actions in realistic conditions, providing validity through psychological and cognitive measurements.

Hybrid instruments have also combined surveys with interviews to enhance quantitative results with qualitative insights. Such an approach is the instrument to measure the cybersecurity awareness and cyber behaviors of college students proposed in \cite{Berry2023}, which utilizes a custom-built survey and interviews to evaluate cognitive and behavioral vulnerabilities, such as password reuse, lack of data backup, and the tendency to prioritize convenience over security. As a result, it offers a more descriptive view of vulnerability, and more specifically around phishing susceptibility, identity theft, and password misuse.

\section{Analysis and Discussion of SLR Results}
\label{sec:Results}
\subsection{Overall Results}
To evaluate the extent to which existing HVA approaches address the multidimensional nature of human cyber vulnerabilities, each identified method, model, or instrument was analyzed and mapped against the vulnerability domains and moderator groups of the Cybersecurity Human Factor Taxonomy designed in this study, to determine which human factors and indicators it explicitly assesses or operationalizes. Table \ref{tab:tabl3} below presents the complete list of studies identified through the PRISMA methodology as eligible for inclusion in this SLR, ordered by year of publication (ascending), along with their mapping to the proposed taxonomy. For each study, the table reports whether it has been classified as a Method (Me), Model (Mo), or Instrument (I).

\begin{figure}[H]
    \centering
    \includegraphics[width=0.9\linewidth]{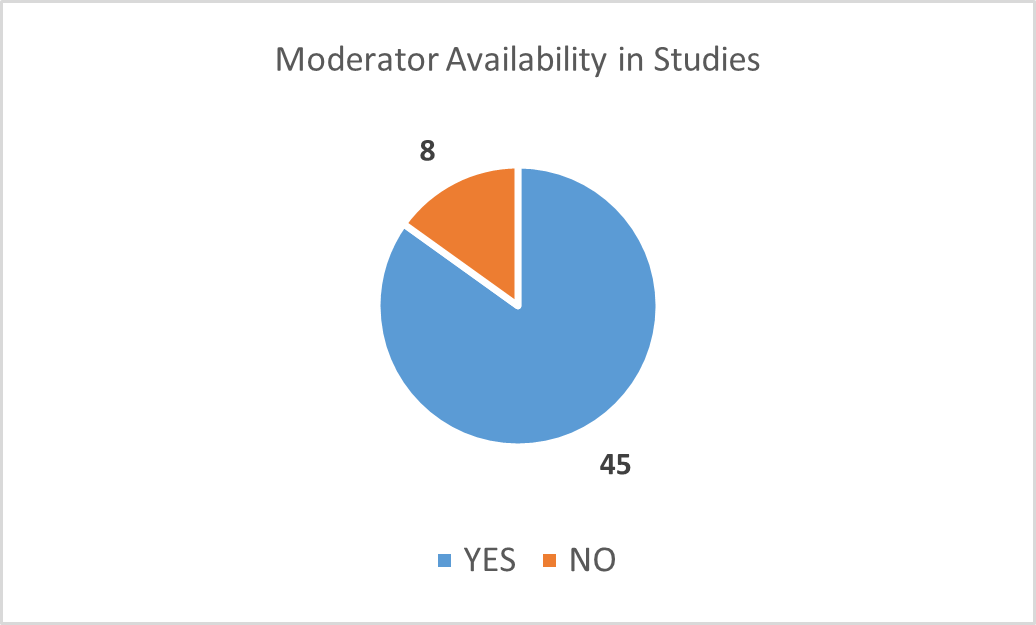}
    \caption{Moderator Availability in Studies}
    \label{fig:fig4}
\end{figure}

The specific human factors and indicators addressed by each study have been recorded in the corresponding vulnerability and moderator columns. All studies included in the SLR addressed at least one vulnerability domain, while 45 out of the 53 included HVA approaches (85\%) considered moderator groups as well, as can be seen in Figure \ref{fig:fig4}. It is worth noting that, due to high variability across the studies regarding the terminology used to describe the human factors or indicators addressed, a conceptual analysis was performed in parallel in order to identify the underlying factors and indicators considered and ensure consistent mapping across studies. For instance, there were studies reporting they addressed "digital literacy", which in some cases referred to Cybersecurity awareness while in others it referred to Computer expertise.

In addition to taxonomy mapping, further analytical dimensions were introduced in order to further characterize how the studies included in the SLR conceptualize and operationalize human cyber vulnerability, namely: i) threat relevance, ii) assessment or measurement approach, iii) vulnerability propagation, iv) vulnerability modelling approach. The threat relevance dimension classifies studies based on whether HVA is performed in light of Unintentional threats (U), Intentional threats (I), or both (U/I), allowing the review to evaluate the extent to which existing HVA approaches consider malicious insider behavior alongside accidental human vulnerabilities. The assessment or measurement approach dimension investigates how human vulnerabilities are practically assessed or measured accross the reviewed studies, such as through self-reported instruments (i.e., surveys), experimental or simulation-based approaches, observation or monitoring procedures for capturing human behavior, machine-learning based models for identifyinf patterns and vulnerabilities, analytical or quantitative models to produce statistical results, expert-based or qualitative approaches, conceptual or theortical methodologies, gamified or interactive methods through which vulnerabilities may surface, and hybrid approaches combining methods.

Furthermore, the vulnerability propagation dimension identifies whether and how a study conceptualizes the interaction or spread of vulnerability across different levels as follows: i) intra-individual refers to cascading interactions between vulnerability-related factors within the same individual, where one condition may influence or amplify another; ii) inter-individual refers to the transmission or amplification of vulnerability between individuals through social interaction, trust relationships, communication patterns, or shared digital behavior, iii) systemic refers to the spread and reinforcement of vulnerabilities through broader socio-technical structures, organizational processes, operational dependencies, governance conditions, or collective behavioral dynamics. 

The vulnerability modelling dimension captures the extent to which vulnerability is conceptualized as fixed, context-dependent, or continuously evolving over time, as follows: i) static conceptualizes vulnerability as a relatively fixed or snapshot-based condition assessed at a specific point in time, typically through one-time measurements or stable profiles; ii) semi-dynamic incorporates contextual or fluctuating conditions, such as stress, fatigue, or situational influences, while still relying primarily on static or cross-sectional assessment approaches; iii) dynamic vulnerability modelling conceptualizes vulnerability as an evolving condition that changes over time according to user behavior, contextual influences, environmental conditions, or system interactions, often incorporating continuous monitoring, adaptive profiling, temporal analysis, or predictive assessment mechanisms.

\subsection{Detailed Results}

A thorough quantitative analysis was performed  on the included studies, using as a basis their mapping to the proposed taxonomy, as depicted in Table \ref{tab:tabl3}, aiming to unravel the aspects of human vulnerability that remain underrepresented among proposed approaches. The following sections provide the detailed results of this analysis, aiming to shed light upon the most commonly considered human factors and moderator indicators as well as the most frequently adopted combinations, assessment or measurement approaches, and the ways vulnerability is modelled and approached.

\subsubsection{Vulnerability Domains \& Human Factors}

\begin{figure}[H]
    \centering
    \includegraphics[width=0.9\linewidth]{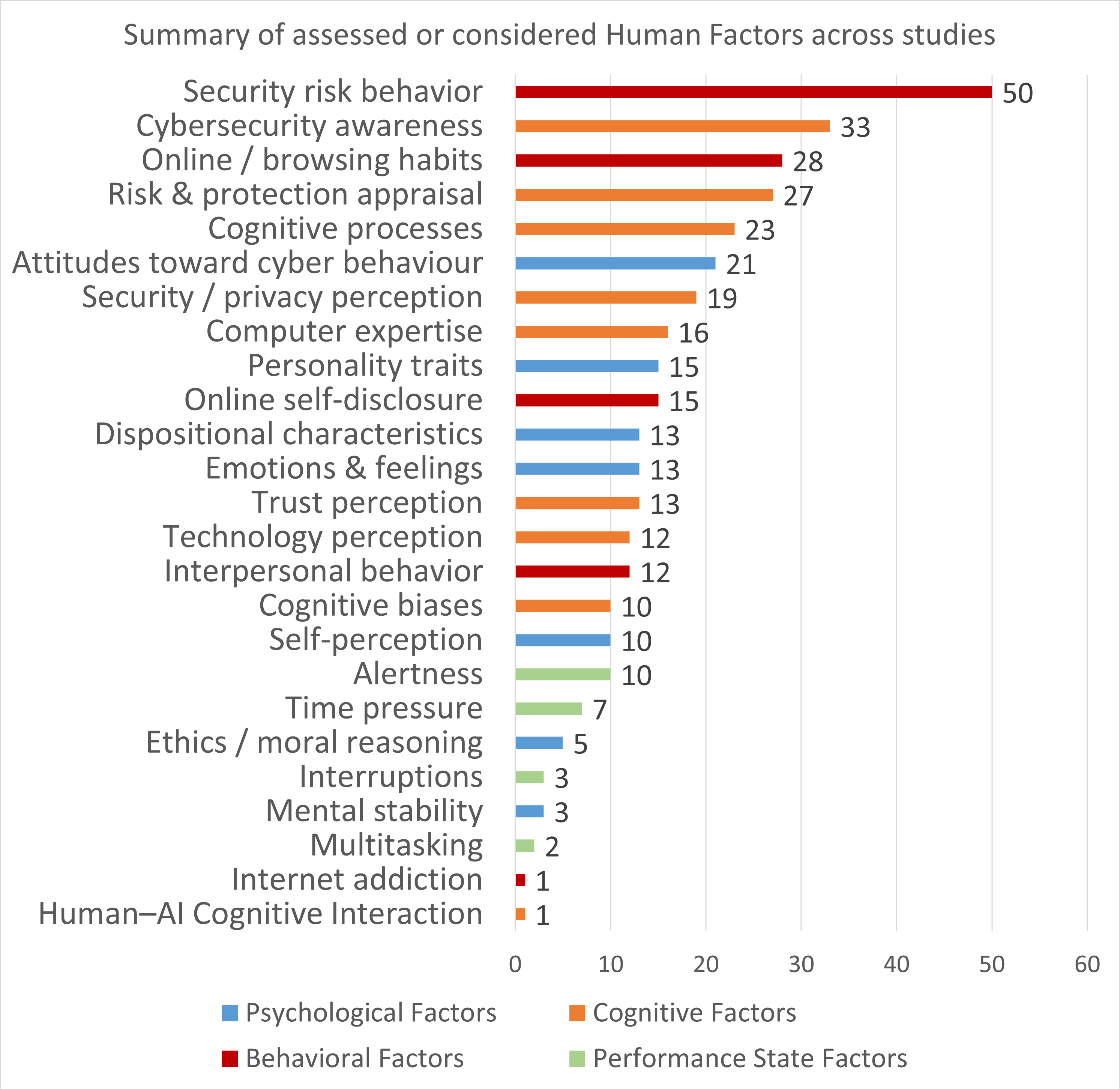}
    \caption{Summary of assessed or considered Human Factors across studies}
    \label{fig:fig9}
\end{figure}

\paragraph{Human Factors considered across Studies}
Figure \ref{fig:fig9} illustrates the overall distribution of all human factors assessed in the reviewed studies, highlighting the frequency in which each factor is considered. The most noticeable studied factor is the Security Risk Behavior behavioral factor, with a high count of 50, indicating a primary focus in the literature among all factors considered. This is followed by the Cybersecurity Awareness and the Risk \& Protection Appraisal cognitive factors, and the Online/browsing Habits behavioral factor, all of which are examined in a considerable number of studies.

Less commonly evaluated factors include among others the Cognitive Biases cognitive factor, the Self-perception psychological factor, as well as the Alertness and Time-pressure performance state factors, with 10 occurences each. Among the least represented factors are Ethics / Morals Reasoning and Mental Stability psychological factors alongside Interruptions and Multitasking performance state factors, appearing only in a limited number of studies. Notably, Internet Addiction and Human-AI Cognitive Interaction behavioral factors, occurring only once, are the most underrepresented factors.

In general, the figure shows that current studies put more emphasis on behavioral and cognitive factors than on psychological and performance state factors. This pattern suggests that current HVA approaches tend to operationalize vulnerability primarily through observable or measurable constructs rather than through more intrinsic or situational aspects. Delving further into the distribution of factors within each vulnerability domain, it is revealed that the dominance of behavioral and cognitive factors is driven by a limited subset of factors assessed within each domain rather than by a balanced consideration of all relevant human vulnerability mechanisms. More complex cognitive mechanisms, such as biases and human-AI interaction, remain underrepresented despite their relevance in modern cyber environments while there lacks a diverse set of behavioral factors capturing different forms of user interaction and exposure to cyber threats.

In addition, the contribution of the psychological domain is concentrated around perception- and attitude-related aspects, while deeper mechanisms related to ethical judgment and psychological stability remain largely unexplored in current assessment approaches, indicating also limited consideration of factors that may be indicative of potential insider threats. Furthermore, situationl aspects such as the Multitasking and Interruptions factors are marginally considered, suggesting a narrow operationalization of performance state factors in current research even though such factors can render an individual more or less vulnerable from one day to the next.

\begin{figure}[H]
    \centering
    \includegraphics[width=0.9\linewidth]{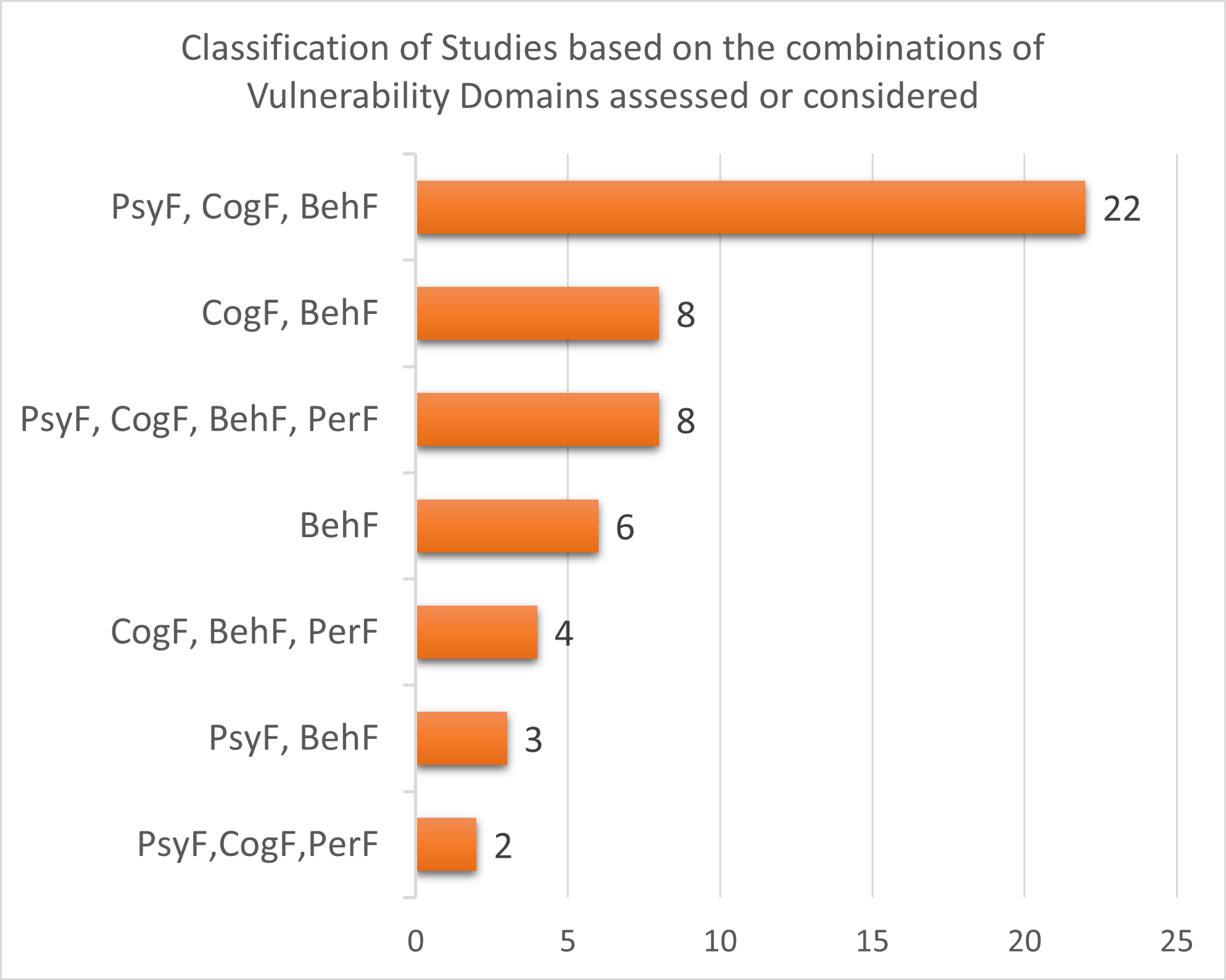}
    \caption{Classification of Studies based on the combinations of Vulnerability Domains assessed or considered}
    \label{fig:fig11}
\end{figure}

\paragraph{Most appeared combinations of Vulnerability Domains \& Human Factors}
Figure \ref{fig:fig11} presents the classification of studies based on combinations of vulnerability domains and how frequently these appear. 

\clearpage
\begin{table*}[!t]
    \centering
    \caption{Mapping of Included Studies to the Cybersecurity Human Factor Taxonomy}
    \label{tab:tabl3}
    \makebox[\textwidth][c]{
    \scriptsize
    \renewcommand{\arraystretch}{1.2}
    \setlength{\tabcolsep}{0.5pt}

    \begin{tabular}{|c|c|*{4}{c|}*{4}{c|}*{5}{c|}}
    \hline

    \multirow{2}{*}{\textbf{Ref.}} & \multirow{2}{*}{\textbf{Categ.}} & \multicolumn{4}{c|}{\textbf{Vulnerability Domains}} & \multicolumn{4}{c|}{\textbf{Moderator Groups}} & \multirow{2}{*}{\textbf{\makecell{Threat \\ Rel.}}} & \multirow{2}{*}{\textbf{\makecell{Ass/Meas. \\ Approach}}} & \multirow{2}{*}{\textbf{\makecell{Vuln. \\ Prop}}} & \multirow{2}{*}{\textbf{\makecell{Vuln. \\ Mod}}} & \multirow{2}{*}{\textbf{\makecell{Pub. \\ Year}}}\\ \cline{3-10} 
    
    & & \makecell{PsyF} & \makecell{CogF} & \makecell{BehF} & \makecell{PerF} & Dem & \makecell{CTr} & \makecell{Cexp} & \makecell{SCE} & & & & & \\ \hline
    
    \cite{Parsons2014} & \makecell{I \\ (RQ3)} & \makecell{Personality \\ traits, \\ Attitudes toward \\ cyber behaviour} & \makecell{Cybersecurity \\ awareness} & \makecell{Security risk \\ behavior, Online \\ self-disclosure, \\ Online / browsing \\ habits} & - & \makecell {Age, \\ Gender} & \makecell{Awareness \\ raising, \\ Formal \\ training} & - & \makecell {Culture, \\ Norms} & U & \makecell{Self- \\ Reported} & \makecell{Intra -\\ individual} & \makecell{Static} & 2014\\ \hline

    \cite{Egelman2015} & \makecell{I \\ (RQ3)} & \makecell{Attitudes toward \\ cyber behaviour} & \makecell{Cybersecurity \\ awareness, \\ Security \\ / privacy perception} & \makecell{Security risk \\ behavior, Online / \\ browsing habits} & - & \makecell{Age, \\ Gender, \\ Education \\ level} & \makecell{Awareness \\ raising} & - & \makecell{Economic \\ stability} & U & \makecell{Self- \\ Reported} & \makecell{None} & \makecell{Static} & 2015\\ \hline

    \cite{PachecoB} & \makecell{Mo \\ (RQ2)} & \makecell{Personality \\ traits, \\ Dispositional \\ characteristics} & \makecell{Cognitive biases, \\ Technology \\ perception,  \\ Security \\ / privacy \\ perception} & \makecell{Security risk \\ behavior, \\ Interpersonal \\ behavior, Online \\ self-disclosure, \\ Online / browsing \\ habits} & \makecell{Time \\ pressure,\\ Multitasking \\ Interruptions} & - & - & - & \makecell{Culture, \\ Societal \\ conditions} & U & \makecell{Experimental/ \\ Simulation- \\ Based} & \makecell{None} & \makecell{Semi- \\ dynamic} & 2017\\ \hline

    \cite{Haim2017} & \makecell{Mo \\ (RQ2)} & - & - & \makecell{Security risk \\ behavior} & - & Occupation & - & - & - & I & \makecell{Behavioral \\ Observation / \\ Monitoring} & \makecell{Systemic} & \makecell{Dynamic} & 2017\\ \hline

    \cite{Alissa2021} & \makecell{I \\ (RQ3)} & \makecell{Attitudes toward \\ cyber behaviour} & - & \makecell{Security risk \\ behavior, \\ Interpersonal \\ behavior, Online \\ self-disclosure} & - & - & - & - & - & U & \makecell{Self \\ Reported} & \makecell{None} & \makecell{Static} & 2018\\ \hline

    \cite{Shi2018} & \makecell{Mo \\ (RQ2)} & - & - & \makecell{Security risk \\ behavior, Online / \\ browsing habits} & - & - & - & - & - & I & \makecell{Machine \\ Learning- \\ Based} & \makecell{Intra- \\ individual} & \makecell{Dynamic} & 2018\\ \hline

    \cite{Archibald2018} & \makecell{Mo \\ (RQ2)} & - & \makecell{Cybersecurity \\ awareness} & \makecell{Security risk \\ behavior, \\ Interpersonal \\ behavior, Online / \\ browsing habits} & \makecell {Time \\ pressure} & - & \makecell{Awareness \\ raising, \\ Formal \\ training} & - & \makecell{Governance \\ environment} & U & \makecell{Hybrid \\ (Experiment, \\ Interviews)} & \makecell{None} & \makecell{Static} & 2018\\ \hline

    \cite{Cullen2018} & \makecell{Me \\ (RQ1)} & \makecell{Personality \\ traits} & \makecell{Cognitive \\ processes, Risk \& \\ protection appraisal} & \makecell{Security risk \\ behavior} & - & - & \makecell{Awareness \\ raising} & - & - & U & \makecell{Hybrid \\ (Survey, \\ Experiment)} & \makecell{Intra- \\ individual} & \makecell{Semi- \\ dynamic} & 2018\\ \hline

    \cite{Miehling2018} & \makecell{Me \\ (RQ1)} & - & \makecell{Risk \& protection \\ appraisal} & \makecell{Security risk \\ behavior} & - & - & - & - & - & U & \makecell{Experimental/ \\ Simulation- \\ Based} & \makecell{None} & \makecell{Dynamic} & 2018\\ \hline

    \cite{Modic2018} & \makecell{I \\ (RQ3)} & \makecell{Personality traits, \\ Dispositional \\ characteristics} & \makecell{Cognitive \\ processes, Risk \& \\ protection appraisal} & \makecell{Security risk \\ behavior, \\ Interpersonal \\ behavior} & - & \makecell{Age, \\ Gender, \\ Education \\ level, \\ Occupation} & - & \makecell{Prior \\ detection \\ (success/ \\ failure), \\ Observational \\ exposure} & \makecell{Culture, \\ Norms} & U & \makecell{Self- \\ reported} & \makecell{Intra- \\ individual} & \makecell{Semi- \\ dynamic} & 2018\\ \hline

     \cite{Ani2019} & \makecell{Mo \\ (RQ2)} & \makecell{Self-perception, \\ Attitudes toward \\ cyber behaviour} & \makecell{Cognitive \\ processes, \\ Cybersecurity \\ awareness, \\ Computer expertise, \\ Risk \& protection \\ appraisal, Security / \\ privacy perception} & \makecell{Security risk \\ behavior} & Alertness & \makecell{Age, \\ Gender, \\ Education \\ level, \\ Occupation} & \makecell{Awareness \\ raising, \\ Formal \\ training, \\ Certification} & \makecell{Prior \\ detection \\ (success/ \\ failure)} & \makecell{Culture, \\ Norms, \\ Governance \\ environment} & U & \makecell{Self- \\ reported} & \makecell{Systemic} & \makecell{Semi- \\ dynamic} & 2018\\ \hline

     \cite{Jiang2018} & \makecell{Mo \\ (RQ2)} & - & - & \makecell{Security risk \\ behavior, Online / \\ browsing habits} & - & Occupation & - & - & - & I & \makecell{Behavioral \\ Observation / \\ Monitoring} & \makecell{Intra- \\ Individual \\ \& Systemic} & \makecell{Dynamic} & 2018\\ \hline

     \cite{Alohali2018} & \makecell{Me \\ (RQ1)} & \makecell{Personality \\ traits, \\ Dispositional \\ characteristics, \\ Attitudes toward \\ cyber behaviour} & \makecell{Cybersecurity \\ awareness, \\ Computer expertise, \\ Technology \\ perception, Risk \& \\ protection appraisal, \\ Security / privacy \\ perception} & \makecell{Security risk \\ behavior, Online \\ self-disclosure, \\ Online / browsing \\ habits} & - & \makecell{Age, \\ Gender} & \makecell{Awareness \\ raising} & \makecell{Prior \\ victimization, \\ Prior \\ detection \\ (success/ \\ failure)} & \makecell{Societal \\ conditions, \\ Economic \\ stability, \\ Legal \\ status} & U & \makecell{Hybrid \\ (Behavioral \\ Monitoring / \\ Analysis \& \\ Survey)} & \makecell{Intra- \\ Individual \\ \& Systemic} & \makecell{Dynamic} & 2018\\ \hline

     \cite{Kim2019} & \makecell{Mo \\ (RQ2)} & \makecell{Personality \\ traits} & - & \makecell{Security risk \\ behavior, Online / \\ browsing habits} & - & \makecell{Occupation} & - & - & - & I & \makecell{Machine \\ Learning- \\ Based} & \makecell{Inter- \\ individual} & \makecell{Dynamic} & 2019\\ \hline

     \cite{Shakela2019} & \makecell{Mo \\ (RQ2)} & - & \makecell{Cybersecurity \\ awareness, Risk \& \\ protection appraisal} & \makecell{Security risk \\ behavior, Online \\ self-disclosure} & - & - & - & \makecell{Prior \\ detection \\ (success/ \\ failure), \\ Observational \\ exposure} & - & U & \makecell{Experimental/ \\ Simulation- \\ Based} & \makecell{Intra- \\ Individual \\ \& Systemic} & \makecell{Semi- \\ dynamic} & 2019\\ \hline

     \cite{Alhosani2019} & \makecell{Mo \\ (RQ2)} & \makecell{Self-perception} & \makecell{Cybersecurity \\ awareness, \\ Computer expertise} & \makecell{Security risk \\ behavior} & - & - & \makecell{Awareness \\ raising, Formal \\ training} & \makecell{Defence \\ action \\ adoption} & - & U & \makecell{Behavioral \\ Observation / \\ Monitoring} & \makecell{Intra- \\ Individual \\ \& Systemic} & \makecell{Semi- \\ dynamic} & 2019\\ \hline

    \end{tabular}
    }
\end{table*}

\clearpage

\begin{table*}[!t]
    \centering
    \makebox[\textwidth][c]{
    \scriptsize
    \renewcommand{\arraystretch}{1.2}
    \setlength{\tabcolsep}{0.5pt}

    \begin{tabular}{|c|c|*{4}{c|}*{4}{c|}*{5}{c|}}
    \hline

    \multirow{2}{*}{\textbf{Ref.}} & \multirow{2}{*}{\textbf{Categ.}} & \multicolumn{4}{c|}{\textbf{Vulnerability Domains}} & \multicolumn{4}{c|}{\textbf{Moderator Groups}} & \multirow{2}{*}{\textbf{\makecell{Threat \\ Rel.}}} & \multirow{2}{*}{\textbf{\makecell{Ass/Meas. \\ Approach}}} & \multirow{2}{*}{\textbf{\makecell{Vuln. \\ Prop}}} & \multirow{2}{*}{\textbf{\makecell{Vuln. \\ Mod}}} & \multirow{2}{*}{\textbf{\makecell{Pub. \\ Year}}}\\ \cline{3-10} 
    
    & & \makecell{PsyF} & \makecell{CogF} & \makecell{BehF} & \makecell{PerF} & Dem & \makecell{CTr} & \makecell{Cexp} & \makecell{SCE} & & & & & \\ \hline

     \cite{Astakhova2019} & \makecell{Mo \\ (RQ2)} & \makecell{Emotions \& \\ Feelings} & - & \makecell{Security risk \\ behavior} & - & - & - & - & - & U / I & \makecell{Behavioral \\ Observation / \\ Monitoring} & \makecell{None} & \makecell{Dynamic} & 2019\\ \hline

     \cite{Velki2019} & \makecell{I \\ (RQ3)} & - & \makecell{Cybersecurity \\ awareness, Risk \& \\ protection appraisal, \\ Security / privacy \\ perception} & \makecell{Security risk \\ behavior, Online \\ self-disclosure, \\ Online / browsing \\ habits} & - & \makecell{Age, \\ Gender, \\ Education \\ level} & - & - & \makecell{Culture, \\ Norms} & U & \makecell{Hybrid \\ (Survey \\ \& Experiment)} & \makecell{Intra- \\ individual} & \makecell{Semi- \\ dynamic} & 2019\\ \hline

     \cite{Eftimie2020} & \makecell{Me \\ (RQ1)} & \makecell{Personality \\ traits, \\ Emotions \& \\ Feelings, \\ Dispositional \\ characteristics, \\ Ethics / moral \\ reasoning, Mental \\ stability} & \makecell{Cybersecurity \\ awareness, \\ Computer expertise, \\ Risk \& protection \\ appraisal, Security / \\ privacy perception} & \makecell{Security risk \\ behavior} & \makecell{Time \\ pressure} & \makecell{Occupation} & \makecell{Awareness \\ raising, \\ Formal \\ training} & \makecell{Defense \\ action \\ adoption} & \makecell{Norms, \\ Values, \\ Legal \\ status, \\ Intergroup \\ behavior} & I & \makecell{Hybrid \\ (Behavioral \\ Monitoring / \\ Analysis \& \\ Survey)} & \makecell{Intra- \& Inter- \\ Individual \& \\ Systemic} & \makecell{Dynamic} & 2020\\ \hline

     \cite{Malek2020} & \makecell{Me \\ (RQ1)} & - & - & \makecell{Security risk \\ behavior} & - & - & - & - & - & I & \makecell{Behavioral \\ Observation / \\ Monitoring} & \makecell{Intra- \\ individual} & \makecell{Dynamic} & 2020\\ \hline

     \cite{Balaram2020} & \makecell{Mo \\ (RQ2)} & - & - & \makecell{Security risk \\ behavior, Online / \\ browsing habits} & - & \makecell{Occupation} & - & - & - & I & \makecell{Machine \\ Learning- \\ Based} & \makecell{Intra- \\ Individual \\ \& Systemic} & \makecell{Dynamic} & 2020\\ \hline

     \cite{Bitton2020} & \makecell{Mo \\ (RQ2)} & \makecell{Attitudes toward \\ cyber behaviour} & \makecell{Cognitive \\ processes, \\ Cognitive biases, \\ Cybersecurity \\ awareness, \\ Computer expertise, \\ Trust perception, \\ Technology \\ perception, Risk \& \\ protection appraisal, \\ Security / privacy \\ perception} & \makecell{Security risk \\ behavior, Online \\ self-disclosure, \\ Online / browsing \\ habits} & - & \makecell{Age, \\ Gender, \\ Education \\ level, \\ Occupation} & - & \makecell{Prior \\ detection \\ (success/ \\ failure), \\ Observational \\ exposure} & - & U & \makecell{Hybrid \\ (Survey \\ \& Behavioral \\ Monitoring \& \\ Experiment)} & \makecell{Intra- \\ individual} & \makecell{Dynamic} & 2020\\ \hline

     \cite{Perrotin2020} & \makecell{Mo \\ (RQ2)} & \makecell{Personality traits, \\ Emotions \& \\ feelings, \\ Dispositional \\ characteristics, \\ Ethics / moral \\ reasoning, Mental \\ stability, Self- \\ perception, \\ Attitudes toward \\ cyber behaviour} & \makecell{Cognitive \\ processes, \\ Cognitive biases, \\ Cybersecurity \\ awareness, \\ Computer expertise, \\ Trust perception, \\ Risk \& protection \\ appraisal} & \makecell{Security risk \\ behavior, \\ Interpersonal \\ behavior} & Alertness & Occupation & \makecell{Awareness \\ raising} & - & \makecell{Societal \\ conditions, \\ Norms, \\ Values, \\ Governance \\ environment, \\ Intergroup \\ behavior} & U & \makecell{Hybrid \\ (Behavioral \\ Analysis \& \\ Machine \\ Learning- \\ Based)} & \makecell{Intra- \& Inter- \\ Individual \& \\ Systemic} & \makecell{Dynamic} & 2020\\ \hline

     \cite{Pan2021} & \makecell{Mo \\ (RQ2)} & - & \makecell{Risk \& protection \\ appraisal} & \makecell{Security risk \\ behavior, Online / \\ browsing habits} & - & Occupation & - & - & - & I & \makecell{Machine \\ Learning- \\ Based} & \makecell{None} & \makecell{Dynamic} & 2020\\ \hline

     \cite{Gangire2021} & \makecell{I \\ (RQ3)} & \makecell{Emotions \& \\ feelings, \\ Dispositional \\ characteristics, Self- \\ perception, \\ Attitudes toward \\ cyber behaviour} & \makecell{Cognitive \\ processes, \\ Cognitive biases, \\ Cybersecurity \\ awareness, \\ Computer expertise, \\ Technology \\ perception, Risk \& \\ protection appraisal, \\ Security / privacy \\ perception} & \makecell{Security risk \\ behavior, \\ Interpersonal \\ behavior,  Online \\ self-disclosure, \\ Online / browsing \\ habits} & - & \makecell{Age, \\ Gender, \\ Education \\ level, \\ Occupation} & \makecell{Awareness \\ raising} & - & \makecell{Societal \\ conditions, \\ Governance \\ environment, \\ Intergroup \\ behavior} & U & \makecell{Self-Reported \\ (Survey-Based)} & \makecell{Intra- \\ individual} & \makecell{Static} & 2020\\ \hline

     \cite{Schoenherr2021} & \makecell{I \\ (RQ3)} & \makecell{Personality traits, \\ Dispositional \\ characteristics, \\ Attitudes toward \\ cyber behaviour} & \makecell{Cognitive \\ processes, \\ Cognitive biases, \\ Cybersecurity \\ awareness, \\ Computer expertise, \\ Technology \\ perception, Risk \& \\ protection appraisal, \\ Security / privacy \\ perception} & \makecell{Security risk \\ behavior, \\ Interpersonal \\ behavior,  Online \\ self-disclosure, \\ Online / browsing \\ habits} & - & \makecell{Age, \\ Occupation} & \makecell{Awareness \\ raising, \\ Frequency} & \makecell{Defense \\ action \\ adoption} & \makecell{Governance \\ environment, \\ Legal \\ status} & U & \makecell{Self- \\ Reported} & \makecell{Intra- \\ individual} & \makecell{Static} & 2021\\ \hline

     \cite{Alissa2021} & \makecell{I \\ (RQ3)} & - & \makecell{Cognitive \\ processes, \\ Cybersecurity \\ awareness, Trust \\ perception, \\ Risk \& \\ protection appraisal, \\ Security / privacy \\ perception} & \makecell{Security risk \\ behavior, \\ Interpersonal \\ behavior,  Online \\ self-disclosure, \\ Online / browsing \\ habits} & \makecell{Alertness,\\ Time \\ pressure, \\ Multitasking, \\ Interruptions} & - & - & \makecell{Prior \\ victimization, \\ Prior \\ detection \\ (success/ \\ failure), \\ Observational \\ exposure, \\ Defense \\ action \\ adoption} & - & U & \makecell{Gamified / \\ Interactive} & \makecell{Intra- \\ individual} & \makecell{Semi- \\ dynamic} & 2021\\ \hline

    \end{tabular}
    }
\end{table*}

\clearpage

\begin{table*}[!t]
    \centering
    \makebox[\textwidth][c]{
    \scriptsize
    \renewcommand{\arraystretch}{1.2}
    \setlength{\tabcolsep}{0.5pt}

    \begin{tabular}{|c|c|*{4}{c|}*{4}{c|}*{5}{c|}}
    \hline

    \multirow{2}{*}{\textbf{Ref.}} & \multirow{2}{*}{\textbf{Categ.}} & \multicolumn{4}{c|}{\textbf{Vulnerability Domains}} & \multicolumn{4}{c|}{\textbf{Moderator Groups}} & \multirow{2}{*}{\textbf{\makecell{Threat \\ Rel.}}} & \multirow{2}{*}{\textbf{\makecell{Ass/Meas. \\ Approach}}} & \multirow{2}{*}{\textbf{\makecell{Vuln. \\ Prop}}} & \multirow{2}{*}{\textbf{\makecell{Vuln. \\ Mod}}} & \multirow{2}{*}{\textbf{\makecell{Pub. \\ Year}}}\\ \cline{3-10}
    
    & & \makecell{PsyF} & \makecell{CogF} & \makecell{BehF} & \makecell{PerF} & Dem & \makecell{CTr} & \makecell{Cexp} & \makecell{SCE} & & & & & \\ \hline

     \cite{Rengarajan2021} & \makecell{Mo \\ (RQ2)} & - & \makecell{Cybersecurity \\ awareness} & \makecell{Security risk \\ behavior, Online / \\ browsing habits} & - & Occupation & - & - & - & I & \makecell{Behavioral \\ Observation / \\ Monitoring} & \makecell{Intra- \\ individual} & \makecell{Dynamic} & 2021\\ \hline

     \cite{Zhu2021} & \makecell{Mo \\ (RQ2)} & - & \makecell{Risk \& protection  \\ appraisal} & \makecell{Security risk \\ behavior, Online / \\ browsing habits} & - & - & - & - & - & I & \makecell{Machine \\ Learning- \\ Based} & \makecell{None} & \makecell{Dynamic} & 2021\\ \hline

     \cite{Reeves2021} & \makecell{Me \\ (RQ1)} & \makecell{Emotions \& \\ feelings, \\ Dispositional \\ characteristics, \\ Ethics / moral \\ reasoning, Self- \\ perception, \\ Attitudes toward \\ cyber behaviour} & \makecell{Cognitive \\ processes, \\ Cognitive biases, \\ Trust perception, \\ Technology \\ perception} & \makecell{Security risk \\ behavior, Online / \\ browsing habits} & \makecell{Alertness, \\ Interruptions} & - & - & \makecell{Observational \\ exposure} & Values & U & \makecell{Expert-Based / \\ Qualitative} & \makecell{Intra- \\ individual} & \makecell{Semi- \\ dynamic} & 2021\\ \hline

     \cite{Jureviien2021} & \makecell{Me \\ (RQ1)} & \makecell{Personality traits, \\ Emotions \& \\ feelings, \\ Dispositional \\ characteristics, \\ Ethics / moral \\ reasoning, Self- \\ perception, \\ Attitudes toward \\ cyber behaviour} & \makecell{Cognitive \\ processes, \\ Cybersecurity \\ awareness, \\ Computer expertise, \\ Risk \& protection \\ appraisal, Security / \\ privacy perception} & \makecell{Security risk \\ behavior, \\ Interpersonal \\ behavior, Online \\ self-disclosure, \\ Internet addiction, \\ Online / browsing \\ habits} & - & \makecell{Age, \\ Gender, \\ Education \\ level, \\ Occupation} & \makecell{Awareness \\ raising} & - & \makecell{Culture, \\ Societal \\ conditions, \\ Norms, \\ Values, \\ Economic \\ stability} & U & \makecell{Hybrid \\ (Behavioral \\ Monitoring / \\ Analysis \& \\ Survey)} & \makecell{Intra- \\ individual} & \makecell{Semi- \\ dynamic} & 2021\\ \hline

     \cite{Antunes2021} & \makecell{Me \\ (RQ1)} & \makecell{Self-perception, \\ Attitudes toward \\ cyber behaviour} & \makecell{Cognitive \\ processes, \\ Cognitive biases, \\ Cybersecurity \\ awareness, Trust \\ perception, \\ Technology \\ perception, Risk \& \\ protection appraisal, \\ Security / privacy \\ perception} & \makecell{Security risk \\ behavior, \\ Interpersonal \\ behavior, Online \\ self-disclosure, \\ Online / browsing \\ habits} & - & \makecell{Gender} & \makecell{Awareness \\ raising, \\ Formal \\ training} & - & \makecell{Norms, \\ Governance \\ environment} & U & \makecell{Self- \\ Reported} & \makecell{None} & \makecell{Semi- \\ dynamic} & 2021\\ \hline

     \cite{Solomon2022} & \makecell{Mo \\ (RQ2)} & - & \makecell{Computer expertise, \\ Technology \\ perception, Risk \& \\ protection appraisal, \\ Security / privacy \\ perception} & \makecell{Security risk \\ behavior, \\ Interpersonal \\ behavior, Online / \\ browsing habits} & - & \makecell{Age, \\ Gender, \\ Education \\ level, \\ Occupation} & \makecell{Awareness \\ raising} & \makecell{Prior \\ detection \\ (success/ \\ failure)} & - & U & \makecell{Hybrid \\ (Behavioral \\ Monitoring / \\ Analysis \& \\ Survey)} & \makecell{None} & \makecell{Dynamic} & 2022\\ \hline

    \makecell{\cite{Perrotin2022}\\  \\ \cite{Belloir2022}} & \makecell{Mo \\ (RQ2)} & \makecell{Emotions \& \\ feelings, \\ Dispositional \\ characteristics, \\ Mental stability} & \makecell{Cybersecurity \\ awareness, \\ Computer expertise, \\ Risk \& protection \\ appraisal} & - & \makecell{Alertness, \\ Time \\ pressure} & - & - & - & \makecell{Culture, \\ Norms, \\ Governance \\ environment} & U & \makecell{Analytical / \\ Quantitative \\ Models} & \makecell{Intra- \& Inter- \\ Individual \& \\ Systemic} & \makecell{Dynamic} & 2022\\ \hline
    
    \cite{Lin2023} & \makecell{Me \\ (RQ1)} & - & \makecell{Cognitive \\ processes, Risk \& \\ protection appraisal} & \makecell{Security risk \\ behavior} & Alertness & - & - & - & - & U & \makecell{Expert- \\ Based / \\ Qualitative} & \makecell{Systemic} & \makecell{Dynamic} & 2022\\ \hline

    \cite{Berry2023} & \makecell{I \\ (RQ3)} & \makecell{Attitudes toward \\ cyber behaviour} & \makecell{Cybersecurity \\ awareness, Trust \\ perception, Risk \& \\  protection appraisal, \\ Security / privacy \\ perception} & \makecell{Security risk \\ behavior, Online / \\ browsing habits} & Time pressure & \makecell{Age, \\ Gender} & \makecell{Awareness \\ raising, \\ Formal \\ training} & \makecell{Defense \\ action \\ adoption} & \makecell{Societal \\ conditions} & U & \makecell{Hybrid \\ (Survey \& \\ Interviews)} & \makecell{None} & \makecell{Static} & 2023\\ \hline

    \cite{Feyzov2023} & \makecell{Mo \\ (RQ2)} & \makecell{Personality traits, \\ Emotions \& \\ feelings} & \makecell{Cybersecurity \\ awareness, Trust \\ perception} & \makecell{Security risk \\ behavior} & - & Occupation & \makecell{Awareness \\ raising, \\ Formal \\ training} & - & \makecell{Values} & U & \makecell{Expert- \\ Based / \\ Qualitative} & \makecell{Intra- \\ Individual \\ \& Systemic} & \makecell{Dynamic} & 2023\\ \hline

    \cite{Chaipa2023} & \makecell{Me \\ (RQ1)} & \makecell{Ethics / moral \\ reasoning} & \makecell{Cognitive \\ processes, \\ Computer expertise} & \makecell{Security risk \\ behavior} & - & - & - & - & \makecell{Societal \\ conditions} & I & \makecell{Conceptual / \\ Theoretical} & \makecell{Intra- \& Inter- \\ Individual \& \\ Systemic} & \makecell{Dynamic} & 2023\\ \hline

    \cite{Duman2023} & \makecell{Mo \\ (RQ2)} & \makecell{Personality traits, \\ Emotions \& \\ feelings, \\ Dispositional \\ characteristics, \\ Self-perception} & \makecell{Cognitive processes, \\ Cybersecurity \\ awareness, Trust \\ perception} & \makecell{Security risk \\ behavior} & Alertness & \makecell{Age, \\ Gender, \\ Education \\ level} & - & - & - & U/I & \makecell{Analytical / \\ Quantitative \\ Models} & \makecell{Intra- \\ individual} & \makecell{Semi- \\ dynamic} & 2023\\ \hline

    \cite{Shihepo2023} & \makecell{Me \\ (RQ1)} & \makecell{Attitudes toward \\ cyber behaviour} & \makecell{Cognitive \\ processes, \\ Cybersecurity \\ awareness, \\ Computer expertise, \\ Trust perception, \\ Risk \& protection \\ appraisal} & \makecell{Security risk \\ behavior} & - & \makecell{Occupation} & \makecell{Awareness \\ raising, \\ Formal \\ training} & \makecell{Prior \\ victimization, \\ Defense \\ action \\ adoption} & \makecell{Governance \\ environment} & U & \makecell{Conceptual/ \\ Theoretical} & \makecell{None} & \makecell{Dynamic} & 2023\\ \hline

    \end{tabular}
    }
\end{table*}

\clearpage

\begin{table*}[!t]
    \centering
    \makebox[\textwidth][c]{
    \scriptsize
    \renewcommand{\arraystretch}{1.2}
    \setlength{\tabcolsep}{0.5pt}

    \begin{tabular}{|c|c|*{4}{c|}*{4}{c|}*{5}{c|}}
    \hline

    \multirow{2}{*}{\textbf{Ref.}} & \multirow{2}{*}{\textbf{Categ.}} & \multicolumn{4}{c|}{\textbf{Vulnerability Domains}} & \multicolumn{4}{c|}{\textbf{Moderator Groups}} & \multirow{2}{*}{\textbf{\makecell{Threat \\ Rel.}}} & \multirow{2}{*}{\textbf{\makecell{Ass/Meas. \\ Approach}}} & \multirow{2}{*}{\textbf{\makecell{Vuln. \\ Prop}}} & \multirow{2}{*}{\textbf{\makecell{Vuln. \\ Mod}}} & \multirow{2}{*}{\textbf{\makecell{Pub. \\ Year}}}\\ \cline{3-10}
    
    & & \makecell{PsyF} & \makecell{CogF} & \makecell{BehF} & \makecell{PerF} & Dem & \makecell{CTr} & \makecell{Cexp} & \makecell{SCE} & & & & &\\ \hline

    \cite{McHatton2023} & \makecell{Me \\ (RQ1)} & \makecell{Attitudes toward \\ cyber behaviour} & \makecell{Cognitive \\ processes, \\ Cybersecurity \\ awareness, \\ Technology \\ perception, Security \\ / privacy perception} & \makecell{Security risk \\ behavior, Online \\ self-disclosure, \\ Online / browsing \\ habits} & - & \makecell{Age, \\ Education \\ level} & \makecell{Awareness \\ raising} & \makecell{Defense \\ action \\ adoption} & - & U/I & \makecell{Analytical/ \\ Quantitative \\ Models} & \makecell{None} & \makecell{Dynamic} & 2023\\ \hline

    \cite{Kim2023} & \makecell{Me \\ (RQ1)} & \makecell{Personality traits, \\ Dispositional \\ characteristics} & \makecell{Cognitive \\ processes, \\ Cybersecurity \\ awareness, Security \\ / privacy perception} & \makecell{Online / browsing \\ habits} & - & \makecell{Occupation} & \makecell{Awareness \\ raising} & \makecell{Defense \\ action \\ adoption} & - & I & \makecell{Conceptual/ \\ Theoretical} & \makecell{Intra- \\ individual} & \makecell{Semi- \\ dynamic} & 2023\\ \hline

    \cite{Dixit2024} & \makecell{Mo \\ (RQ2)} & - & - & \makecell{Security risk \\ behavior, Online / \\ browsing habits} & - & - & - & - & - & I & \makecell{Machine \\ Learning- \\ Based} & \makecell{None} & \makecell{Dynamic} & 2024\\ \hline

    \cite{Hussain2024} & \makecell{I \\ (RQ3)} & \makecell{Personality traits} & \makecell{Cybersecurity \\ awareness, Trust \\ perception} & \makecell{Security risk \\ behavior, Online / \\ browsing habits} & - & \makecell{Education \\ level} & \makecell{Awareness \\ raising} & - & - & U & \makecell{Hybrid \\ (Survey \& \\ Experiment)} & \makecell{None} & \makecell{Semi- \\ dynamic} & 2024\\ \hline

    \cite{Li2024} & \makecell{Me \\ (RQ1)} & \makecell{Emotions \& \\ feelings, \\ Attitudes toward \\ cyber behaviour} & \makecell{Cognitive \\ processes, \\ Cybersecurity \\ awareness, Risk \& \\ protection appraisal} & - & Alertness & - & \makecell{Awareness \\ raising, \\ Formal \\ training} & \makecell{Observational \\ exposure} & - & U & \makecell{Conceptual/ \\ Theoretical} & \makecell{Intra- \\ individual} & \makecell{Semi- \\ dynamic} & 2024\\ \hline

    \cite{Mitra2024} & \makecell{Me \\ (RQ1)} & \makecell{Personality traits, \\ Emotions \& \\ feelings, \\ Attitudes toward \\ cyber behaviour} & \makecell{Cognitive \\ processes, \\ Cybersecurity \\ awareness, \\ Conmputer expertise, \\ Trust perception, \\ Technology \\ perception, Risk \& \\ protection appraisal} & \makecell{Security risk \\ behavior, \\ Interpersonal \\ behavior} & - & \makecell{Age, \\ Gender, \\ Education \\ level} & \makecell{Awareness \\ raising} & \makecell{Prior \\ victimization} & \makecell{Culture, \\ Societal \\ conditions, \\ Norms} & I & \makecell{Analytical/ \\ Quantitative \\ Models} & \makecell{None} & \makecell{Semi- \\ dynamic} & 2024\\ \hline

    \cite{Wijesinghe2024} & \makecell{Mo \\ (RQ2)} & \makecell{Attitudes toward \\ cyber behaviour} & \makecell{Cognitive \\ processes, \\ Cybersecurity \\ awareness, \\ Conmputer expertise, \\ Security / privacy \\ perception} & \makecell{Security risk \\ behavior, Online / \\ browsing habits} & - & - & \makecell{Awareness \\ raising, \\ Formal \\ training} & \makecell{Defense \\ action \\ adoption} & - & U & \makecell{Analytical/ \\ Quantitative \\ Models} & \makecell{Intra- \\ individual} & \makecell{Dynamic} & 2024\\ \hline

    \cite{Farzand2024} & \makecell{I \\ (RQ3)} & \makecell{Emotions \& \\ feelings, \\ Self-perception, \\ Attitudes toward \\ cyber behaviour} & \makecell{Cognitive \\ processes, Security \\ / privacy perception} & \makecell{Security risk \\ behavior} & - & \makecell{Age, \\ Gender, \\ Occupation} & \makecell{Awareness \\ raising} & \makecell{Prior \\ victimization} & Culture & U/I & \makecell{Self- \\ Reported} & \makecell{Intra- \\ individual} & \makecell{Static} & 2024\\ \hline

    \cite{Mocerino2024} & \makecell{Mo \\ (RQ2)} & \makecell{Attitudes toward \\ cyber behaviour} & \makecell{Cognitive \\ processes, \\ Cognitive biases, \\ Cybersecurity \\ awareness, \\ Computer expertise} & \makecell{Security risk \\ behavior} & \makecell{Alertness, \\ Time \\ pressure} & \makecell{Age, \\ Gender, \\ Education \\ level, \\ Occupation} & \makecell{Awareness \\ raising, \\ Formal \\ training} & \makecell{Prior \\ victimization} & - & I & \makecell{Analytical/ \\ Quantitative \\ Models} & \makecell{Intra- \\ individual} & \makecell{Static} & 2024\\ \hline

    \cite{Adnan2025} & \makecell{Mo \\ (RQ2)} & - & \makecell{Cybersecurity \\ awareness, \\ Technology \\ perception, Risk \& \\ protection appraisal} & \makecell{Security risk \\ behavior} & - & \makecell{Age, \\ Gender, \\ Education \\ level, \\ Occupation} & \makecell{Awareness \\ raising, \\ Certification} & \makecell{Defense \\ action \\ adoption} & - & I & \makecell{Analytical/ \\ Quantitative \\ Models} & \makecell{None} & \makecell{Semi- \\ dynamic} & 2025\\ \hline

    \cite{Barath2025} & \makecell{Me \\ (RQ1)} & \makecell{Self-perception, \\ Attitudes toward \\ cyber behaviour} & \makecell{Cognitive \\ processes, \\ Cybersecurity \\ awareness, \\ Technology \\ perception, Risk \& \\ protection appraisal, \\ Human-AI Cognitive \\ Interaction} & \makecell{Security risk \\ behavior, Online \\ self-disclosure} & - & \makecell{Age, \\ Education \\ level} & \makecell{Awareness \\ raising, \\ Formal \\ training} & - & - & U & \makecell{Self- \\ Reported} & \makecell{None} & \makecell{Semi- \\ dynamic} & 2025\\ \hline

    \cite{Duman2025} & \makecell{Mo \\ (RQ2)} & - & \makecell{Cognitive biases, \\ Cybersecurity \\ awareness, Trust \\ perception} & \makecell{Security risk \\ behavior} & Alertness & \makecell{Age, \\ Gender, \\ Education \\ level} & - & - & \makecell{Economic \\ stability} & U & \makecell{Analytical/ \\ Quantitative \\ Models} & \makecell{Intra- \\ individual} & \makecell{Dynamic} & 2025\\ \hline

    \cite{Papatsaroucha2025} & \makecell{Mo \\ (RQ2)} & \makecell{Emotions \& \\ feelings, \\ Dispositional \\ characteristics} & \makecell{Cognitive biases, \\ Cybersecurity \\ awareness, Trust \\ perception, Risk \& \\ protection appraisal, \\ Security / privacy \\ perception} & \makecell{Security risk \\ behavior, Online \\ self-disclosure, \\ Online / browsing \\ habits} & - & - & - & \makecell{Prior \\ detection \\ (success/ \\ failure)} & - & U & \makecell{Experimental/ \\ Simulation- \\ Based} & \makecell{None} & \makecell{Dynamic} & 2025\\ \hline

    \end{tabular}

    }
\end{table*}

\clearpage

Studies were identifying as addressing a particular domain if they assessed at least one human factory of the domain. As can be seen, the most common combination, with 22 studies, includes factors of the psychological, cognitive, and behavioral domains, indicating a general recognition in the literature that human vulnerability emerges from multi-layered interactions rather than from isolated factors and representing a tendency towards adopting a comprehensive approach.

The next most frequent combination, although represented by significantly fewer studies, appears to be cognitive (CogF) and behavioral (BehF) factors, with 8 occurrences. The absence of inclusion of Psychological Factors (PsyF) in this combination further supports the observation that there is a number of studies that tends to prioritize measurable constructs, such as awareness and behavior, over deeper underlying mechanisms. The same number of studies was also identified for the combination including factors of all vulnerability domains, further highlighting the tendency towards multi-domain approaches. Despite the fact that this finding constitutes an important observation, the limited number of studies combining all vulnerability domains, compared to the studies including the PsyF-CoF-BehF combination, suggests that performance-related aspects are partially included alongside the three other domains. This observation further suggests the finding that situational and dynamic conditions are not systematically integrated into vulnerability assessment approaches. Moreover, even though single-domain approaches are less common, it is worth noting that when a single domain is considered this tends to correspond to behavioral factors, appearing in 6 studies.

In addition, a deeper examination reveals that the domains included in multi-domain approaches are not equally represented in terms of the number of factors originating from each domain and included in the assessment. In particular, studies classified under the PsyF–CogF–BehF combination, include a higher number of behavioral factors, ranging between two and four per study, compared to a more limited inclusion of psychological factors, which are often restricted to one or two. Cognitive factors tend to occupy an intermediate position, with approximately two to three factors per study. A similar pattern is observed in studies incorporating all four domains where performance state factors are included but remain limited in number, typically appearing as one or two factors alongside a stronger representation of behavioral and cognitive factors. This adds to the observation that, even in approaches aiming to move towards holistic assessments, situational and dynamic aspects are treated as secondary additions rather than as core components of vulnerability assessment. These observations reinforce the pattern depicted in Figure \ref{fig:fig9}, suggesting that current HVA approaches, even in multi-domain vulnerability contexts, often exhibit structural imbalance, with behavioral factors consistently receiving more emphasis.

\begin{figure}[H]
    \centering
    \includegraphics[width=0.9\linewidth]{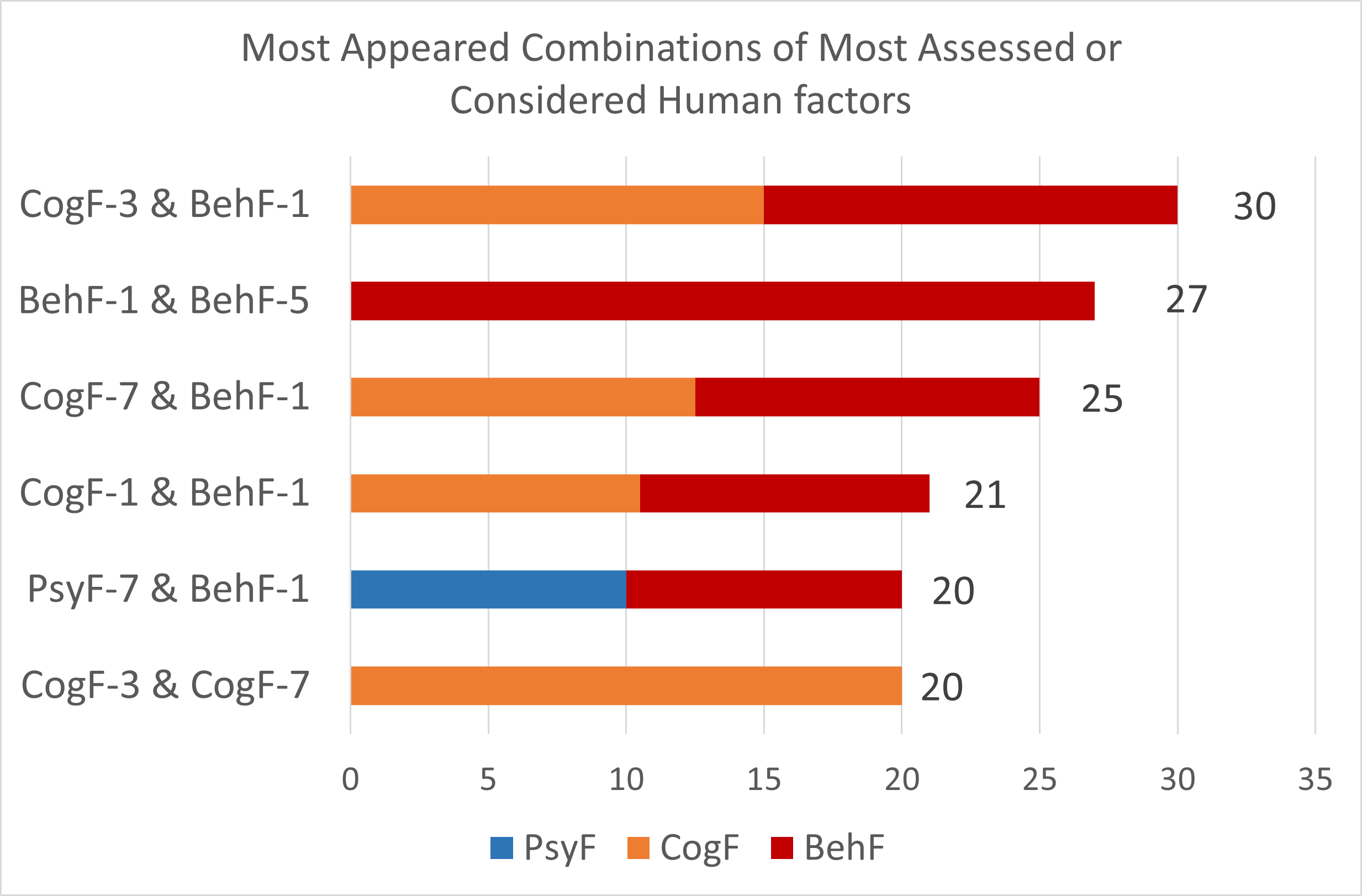}
    \caption{Most Appeared Combinations of Most Assessed or Considered Human factors}
    \label{fig:fig10}
\end{figure}

An additional analysis was performed to identify the most common combinations of the most frequently assessed human factors across the studies, illustrated in Figure \ref{fig:fig10}. The most common combination is between Security Risk Behavior (BehF-1) and Cybersecurity Awareness (CogF-3), with 30 occurences, followed closely by Security Risk Behavior (BehF-1) combined with the users’ Online/browsing Habits (BehF-5) and with Risk \& Protection appraisal (CogF-7), with 27 and 25 occurences, respectively. Less frequent combinations include Security Risk Behavior (BehF-1) with Cognitive Processes (CogF-1) or Attitudes toward Cyber Behavior (PsyF-7), as well as Cybersecurity Awareness (CogF-3) with Risk \& Protection Appraisal (CogF-7). Even though in this comparison these combinations appear less frequently than the most common ones, the number of studies including them remains considerable (20-21 studies), further higlighting the tendency towards combining behavioral and cognitive factors, in particular security risk behavior and risk protection appraisal.

As observed, it is revealed that Security Risk Behavior (BehF-1) is consistently included as the common factor across high-frequency combinations. It is worth noting also that combinations that do not include Security Risk Behavior (BehF-1), such as the pairing of Cybersecurity Awareness (CogF-3) and Risk \& Protection Appraisal (CogF-7), are rarely observed, suggesting that cognitive factors are not commonly assessed independently of behavioral outcomes. Furthermore, the limited inclusion of psychological factors in high-frequency combinations further highlights that psychological mechanisms are less frequently integrated into multi-factor assessment approaches. 

The comparison between domain-level combination and factor-level dominance reveals that current HVA approaches tend to be structurally multi-domain, having realized the multi-facet nature of human vulnerability, but functionally behavior-centric as the focus tends to be on factors related to observable user actions rather than on a balanced integration of psychological, cognitive, and behavioral mechanisms. In this sense, the strong emphasis on behavioral indicators suggests that many existing approaches primarily capture vulnerability at the point of manifestation, i.e., while users have already initiated an engagement in potentially risky or unsafe actions. Even though this provides valuable insights into observed behavior and human susceptibility hotspots in real-time and, potentially, real-world conditions, i.e., while humans operate within digital systems, it may limit the ability of such approaches to proactively identify underlying vulnerability conditions before they cascade into observable security-relevant actions.

\subsubsection{Moderator Groups \& Indicators}
The sections below discuss the indicators considered by the studies (45) that included the assessment or measurement of Moderator Groups. It should be clarified that most of the reviewed studies did not consider Moderator Groups and indicators idependently from Vulnerability Domains and human factors; however, the analysis performed in this study presents the results separately to stay consistent with the proposed taxonomy. After the presentation of these results, this SLR follows an integrated approach to evaluate the coverage of both human factors and indicators of the reviewed studies, aiming to assess their holisticness towards HVA.

\begin{figure}[H]
    \centering
    \includegraphics[width=0.9\linewidth]{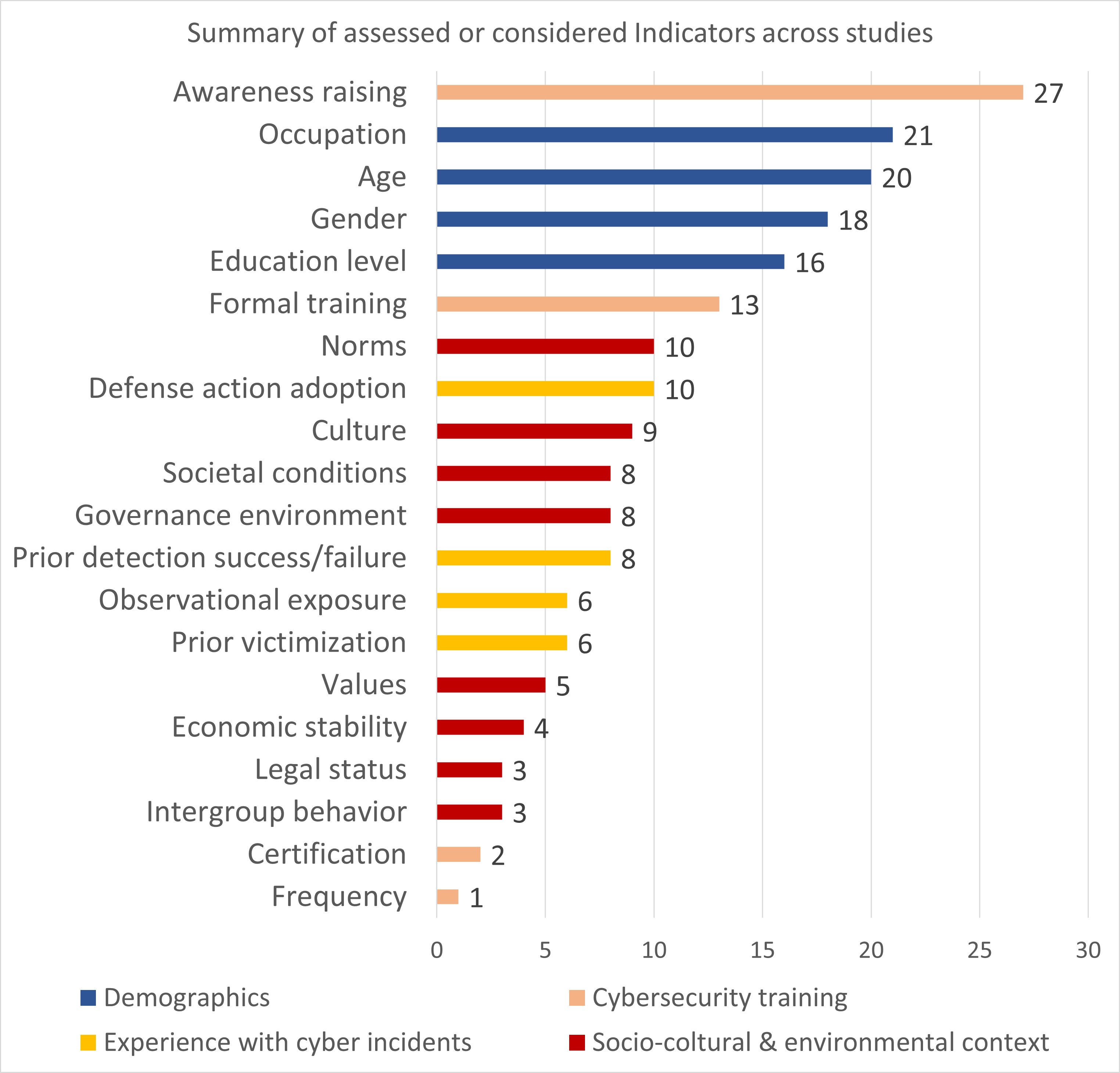}
    \caption{Summary of assessed or considered Indicators across studies}
    \label{fig:fig16}
\end{figure}

\paragraph{Indicators considered across Studies}
Figure \ref{fig:fig16} provides an overview of the frequency of all assessed or considered moderator indicators. As can be seen, Awareness raising, from the Cybersecurity Training moderator group, is the most dominant indicator with 27 occurrences, followed closely by Demographics indicators, such as Occupation (21 occurences), Age (20 occurences), and Gender (18 occurences). At the mid-level of frequency, indicators such as Education level from the Demographics group and Formal Training from the Cybersecurity Training group are observed, along with Norms from the Socio-cultural \& Environmental Context group and Defense Action Adoption from the Experience with Cyber-incidents group. The least represented include indicators such as Economic Stability and Legal Status from the Socio-cultural \& Environmental Context group, and Certification and Frequency of the Cybersecurity Training group.

This distribution reveals that while Awareness raising and Demographics have often been considered in HVA approaches, most Socio-cultural \& Environmental Context indicators remain largely underexplored, despite their potential influence in vulnerabilities related to both unintentional and intentional threats. This distribution further highlights that current approaches tend to rely on observable or easily measurable attributes while more complex contextual conditions are less frequently considered, even though they may influence how vulnerabilities emerge and evolve and are also challenging to foresee. In addition, the dominance of Awareness raising as the most frequently considered indicator suggests that cybersecurity training is often treated as a factor indicative of potential susceptibility without though systematically distinguishing between different forms of training, their frequency, or their effectiveness, which is further supported by the limited presence of Certification and training Frequency indicators.

\begin{figure}[H]
    \centering
    \includegraphics[width=0.9\linewidth]{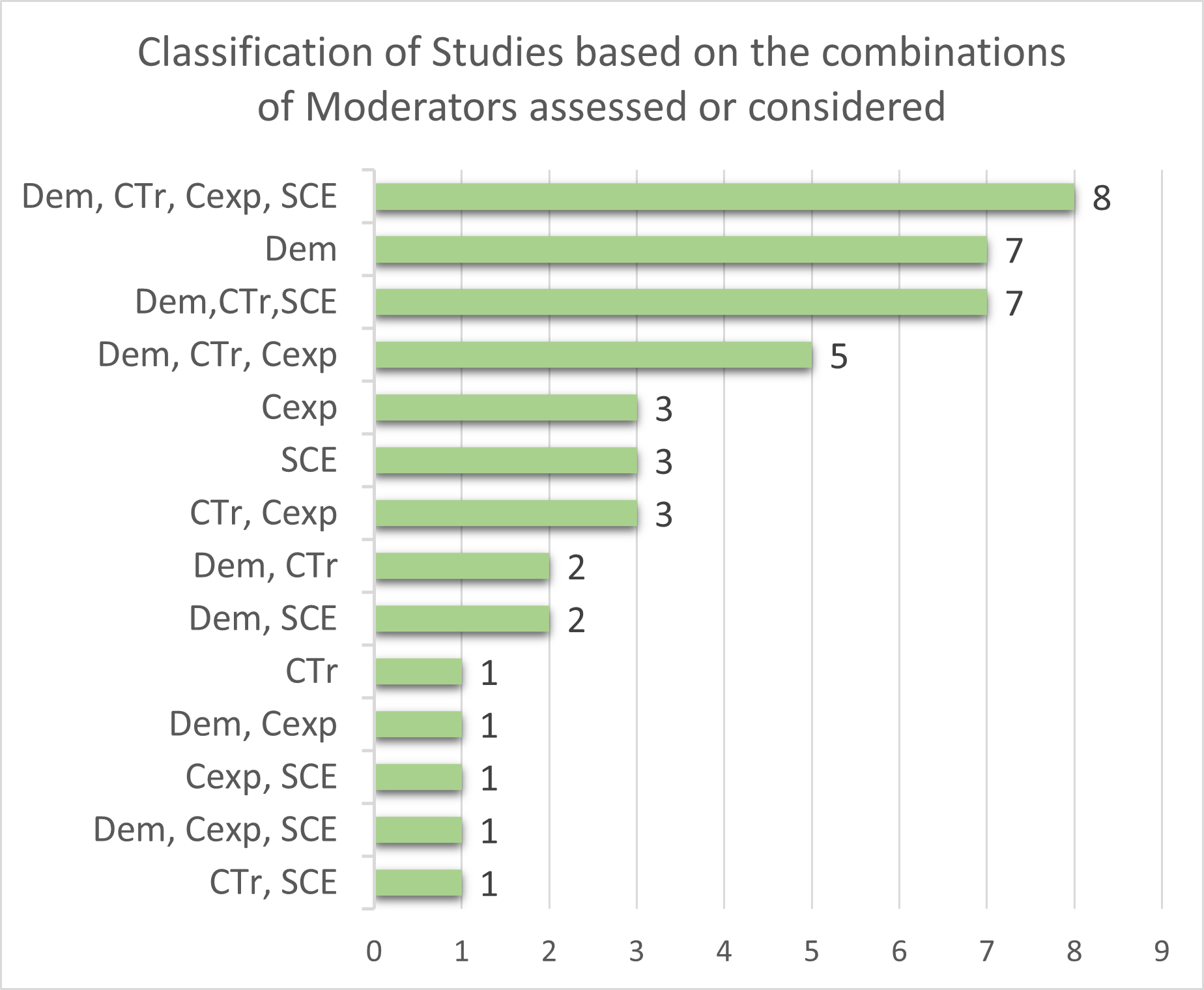}
    \caption{Classification of Studies based on the combinations of Moderator Groups assessed or considered}
    \label{fig:fig18}
\end{figure}

\paragraph{Most Appeared Combinations of Moderator Groups \& Indicators}
Figure \ref{fig:fig18} illustrates the classification of reviewed studies based on the specific combinations of moderator groups they assess or consider. The analysis shows that the most common combination includes all four moderator groups (Dem-CTr-Cexp-SCE, 8 occurences). While this is an important observation, it is followed closely by studies focusing exclusively on a single moderator group, i.e., Demographics with 7 occurences. Notably, the Socio-cultural \& Environmental Context group has been largely considered across the identified combinations, being combined with other groups in 20 studies and assessed as a stand-alone group in 3 studies, bringing the total representation of this group to 23 studies. This observation could be interpreted as contradictory to the findings stemming from Figure \ref{fig:fig16}, but actually indicates that researchers increasingly acknowledge the broader environment as a variable affecting human vulnerability that should be considered in combination with other indicators. 

A deeper examination of the internal composition of moderator group combinations was performed in a similar manner as for the vulnerability domains presented in the previous section. With regard to the latter observation about the inclusion of the Socio-cultural \& Environmental Context group across 20 combinations, it is revealed that the inclusion of multiple moderator categories within a study does not necessarily correspond to a balanced representation of indicators from each group. Emphasis is often placed on easily measurable or commonly available indicators. For instance, studies incorporating all four moderator groups include a higher number of Demographic indicators, while indicators from the Socio-cultural \& Environmental Context group are more often than not limited to one or two aspects. 

Across other combinations, Cybersecurity Training indicators are most frequently represented by Awareness raising, which was anticipated based on Figure \ref{fig:fig16}, while more specific indicators such as Certification (CTr-3) and training Frequency (CTr-4) are rarely included. The Experience with Cyber-incidents group is also selectively considered, most commonly through Defense Action adoption (Cexp-4) or prior Detection success/failure (Cexp-2), rather than through a broader representation of experiential dimensions. Overall, these patterns suggest that current approaches tend to be structurally inclusive but contextually shallow, in the sense that multiple moderator groups may be present, but their role in shaping vulnerability is not equally emphasized or systematically explored.

\begin{figure}[H]
    \centering
    \includegraphics[width=0.9\linewidth]{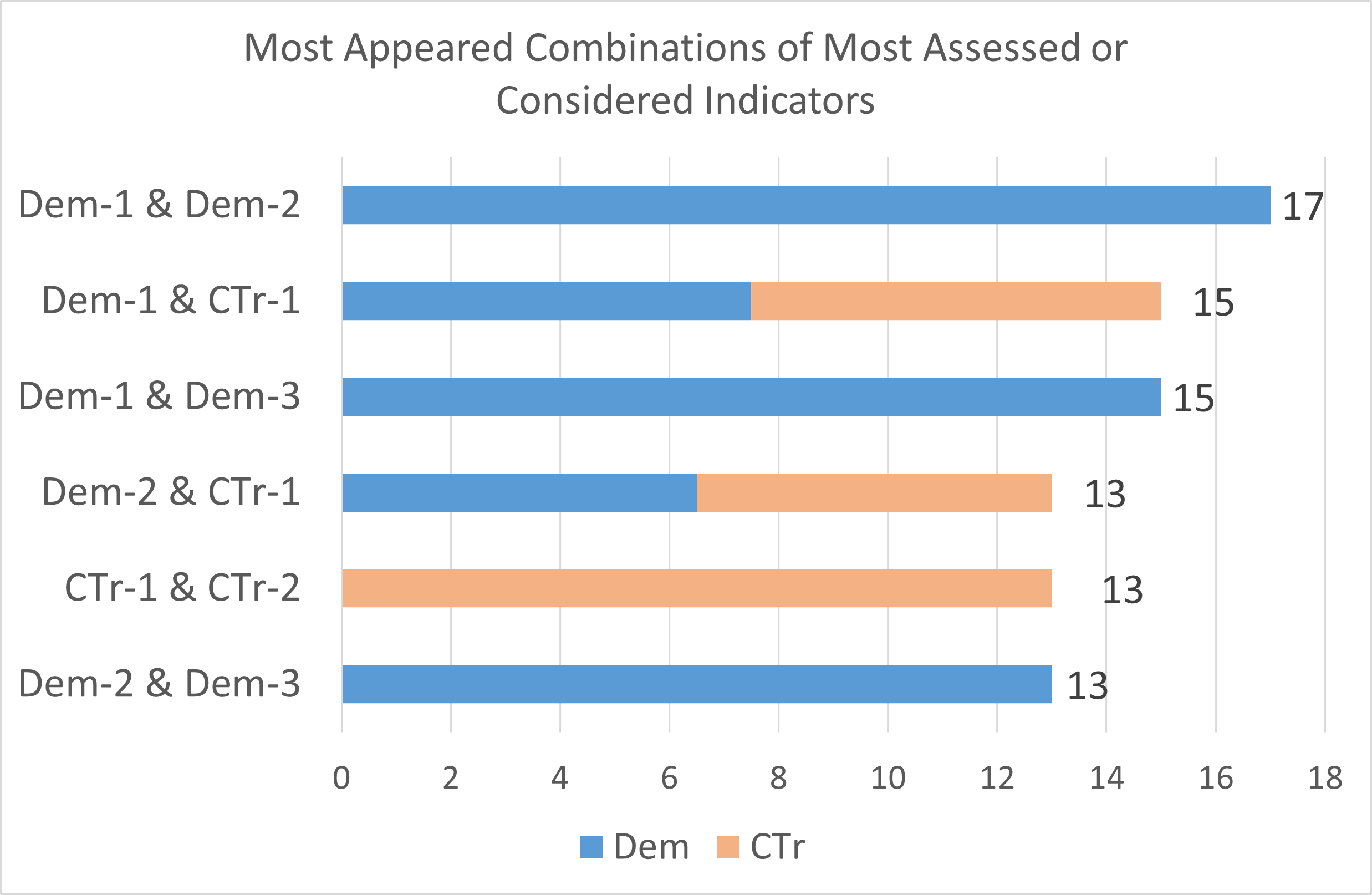}
    \caption{Most appeared Combinations of Most Assessed or Considered Indicators}
    \label{fig:fig17}
\end{figure}

This observation is further reinforced when delving deeper into the most appeared combinations of the most assessed or considered indicators presented in Figure \ref{fig:fig17}, from which indicators of the SCE moderator group are completely absent as they are less frequently assessed, suggesting that comprehensive contextual modeling is not yet systematically adopted. The most common combination is Dem-1-Dem-2 (17 occurrences), representing a tendency to pair the Age and Gender indicators of the Demographics moderator group. This is followed by combinations of Age and Awareness Raising (Dem-1-CTr-1), and further demographic combinations. Other notable combinations include Gender and Awareness Raising (Dem-2-CTr-1), Awareness Raising and Formal Training (CTr-1-CTr-2), and Gender and Education level (Dem-2-Dem-3).  

The frequent combination of Demographic indicators with Awareness Raising (CTr-1) and Formal Training (CTr-2) implies that training-related indicators are often evaluated in conjunction with user profiles, rather than as independent or context-sensitive elements. This observation highlights the tendency for moderator indicators to be employed for user categorization (e.g., age, gender, education), rather than to illustrate how contextual factors dynamically affect vulnerability within the broader operational environment of the user. These findings indicate that although moderator variables are widely recognized in the literature, their incorporation into HVA approaches remains mostly descriptive and user-focused, rather than contextually driven and system-oriented, which may hinder the capacity of existing approaches to effectively capture the influence of complex and evolving real-world conditions on vulnerability. 

\subsubsection{Holisticness}
\begin{figure*}[t]
    \centering
    \includegraphics[width=1.0\textwidth]{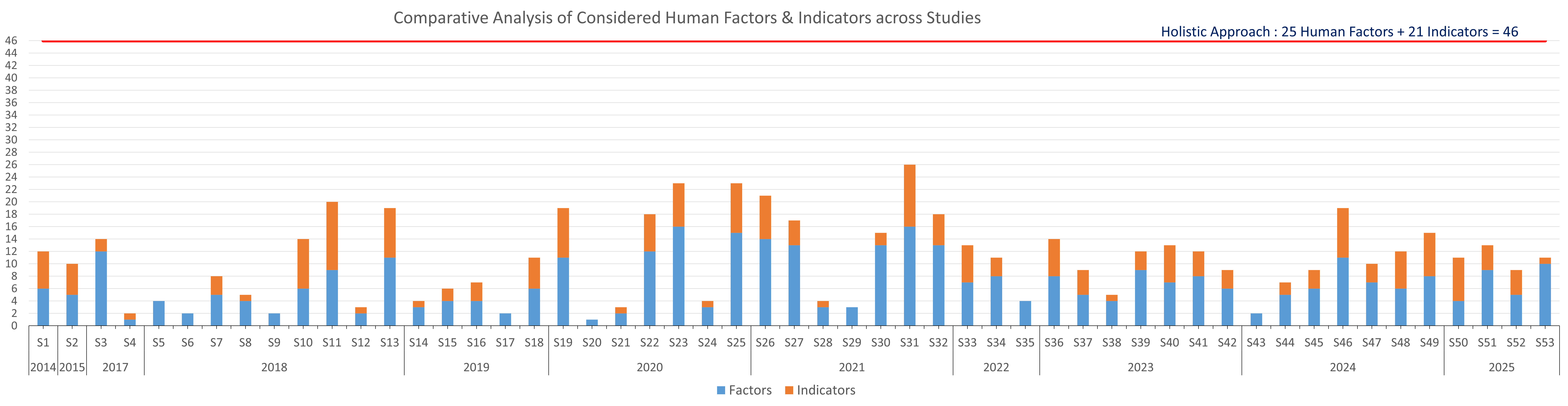}
    \caption{Comparative Analysis of Considered Human Factors \& Indicators across Studies}
    \label{fig:Holistic}
\end{figure*}

An important finding of this SLR refers to the holisticness ratio of the proposed HVA solutions across the reviewed studies. Holisticness is interpreted as the breadth of vulnerability-related dimensions considered by a proposed HVA solution, including both human factors and moderator indicators, as depicted in Figure \ref{fig:Holistic}. Based on the proposed taxonomy, a holistic approach could include up to 46 variables in total, incorporating both the 25 identified human factors and the 21 identified moderator indicators.

As revealed by the quantitative analysis evaluating human factor and indicator coverage, only a limited number of studies has considered a broad range of variables across multiple vulnerability domains and moderator groups. Particularly, the highest number of combined variables identified in a single study was 26 out of the 46 possible variables, resulting to a holisticness ratio of approximately 56.5\%, indicating that even the study with the highest holisticness ratio still considers slightly more than half of the potential vulnerability-related dimensions and contextual indicators considered in this review. In addition, it is revealed that the holisticness ratio does not correspond to balanced representations of factors and indicators. Holisticness has been considered mostly with regard to breadth rather than depth, which remains limited as anticipated by the analysis provided above. Behavioral and cognitive factors are dominating assessment approaches with regard to human factors while contextual moderator variables and dynamic conditions, such as socio-cultural dimensions, remain underrepresented even among the studies with the highest holisticness ratios.

Overall, the findings can be considered rather on the optimistic side, as they suggest that current HVA approaches acknowledge the multi-facet nature of human vulnerability and investigate assessment approaches that aim to account for more than one dimension. Nevertheless, this holisticness remains structurally imbalanced.

\subsubsection{Threat Relevance}

\begin{figure}[H]
    \centering
    \includegraphics[width=0.9\linewidth]{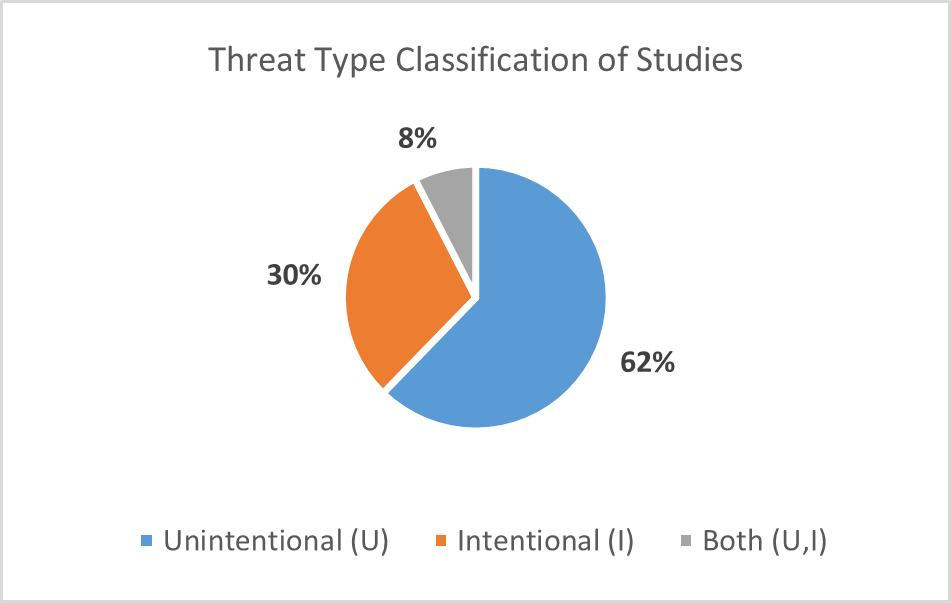}
    \caption{Threat Type Classification of Studies}
    \label{fig:Threat}
\end{figure}

Another important finding of this SLR stems from the the classification of the included studies according to the Threat Type addressed by the methods, models, and instruments they propose, presented in Figure \ref{fig:Threat}. As illustrated, the majority of the studies (62\%) address solely vulnerabilities that may lead to Unintentional Threats, while fewer than half of them (30\%) address vulnerabilities relevant to Intentional Threats. It is important to note that, despite the recognition of human factors and moderator indicators across the literature that are indicative of potential malicious behavior, only 8\% of the proposed solutions address both Unintentional and Intentional Threats. This highlights an important gap in current methodologies and implementations and a significant limitation that needs to be considered for moving towards the development of holistic solutions that realise that humans may threaten systems and data both accidentally and deliberately.

\begin{figure}[H]
    \centering
    \includegraphics[width=0.9\linewidth]{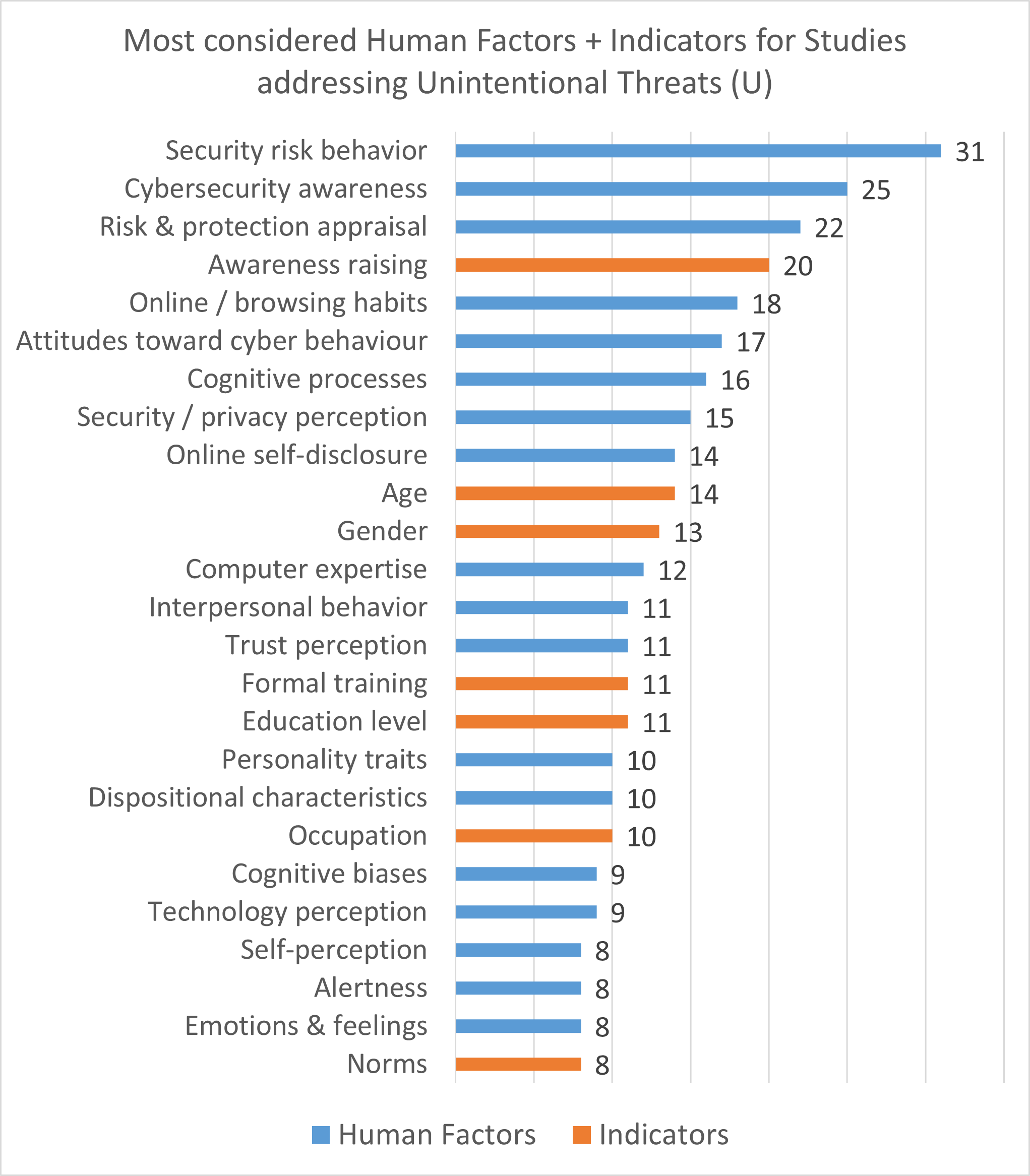}
    \caption{Most considered Human Factors + Indicators for Studies addressing Unintentional Threats (U)}
    \label{fig:U}
\end{figure}

\begin{figure}[H]
    \centering
    \includegraphics[width=0.9\linewidth]{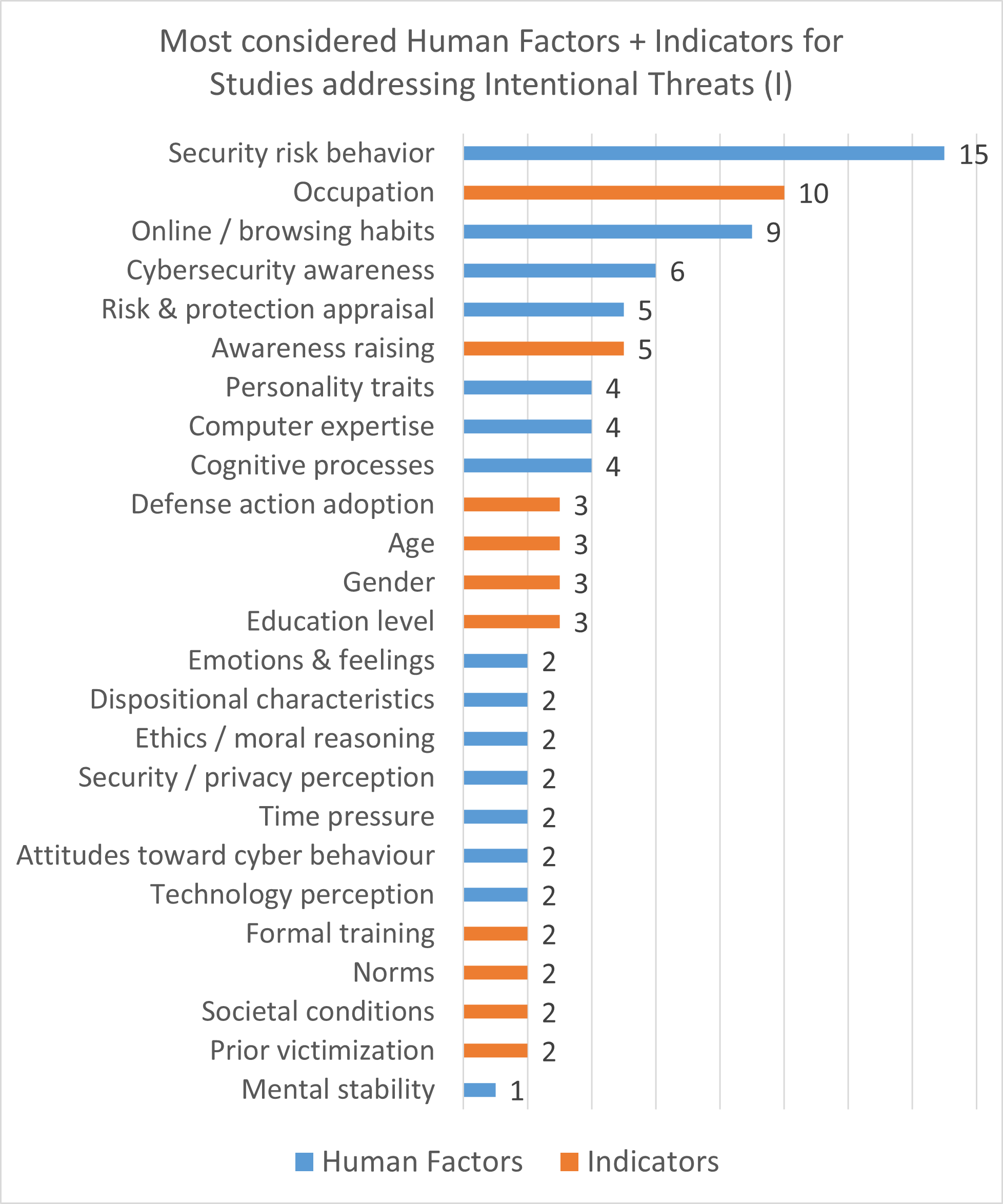}
    \caption{Most considered Human Factors + Indicators for Studies addressing Intentional Threats (I)}
    \label{fig:I}
\end{figure}

\begin{figure}[H]
    \centering
    \includegraphics[width=0.9\linewidth]{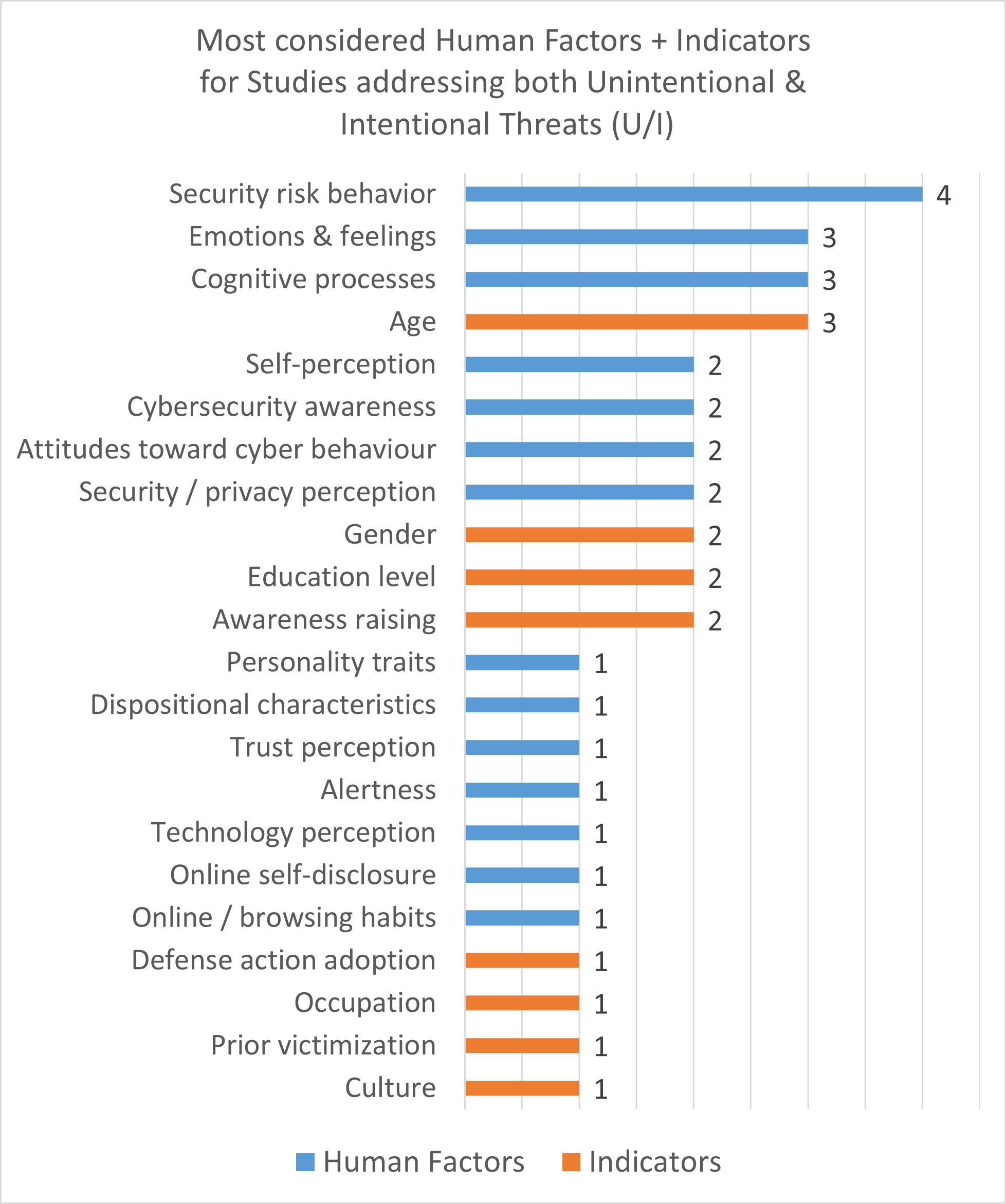}
    \caption{Most considered Human Factors + Indicators for Studies addressing both Unintentional \& Intentional Threats (U-I)}
    \label{fig:UI}
\end{figure}

Furthermore, Figures \ref{fig:U}, \ref{fig:I}, and \ref{fig:UI} provide additional information regarding the human factors and indicators that dominate the assessment with regard to the threat type addressed. Because the number of studies differs across the three threat categories, the interpretation of these figures considers both absolute occurrences and relative prevalence within each category. Within this context, a clear pattern emerges as, across all three classifications, the Security Risk Behavior factor holds the highest rank with the most occurences in suggested approaches, independently of whether the assessed threats are intentional, unintentional, or both.

In studies addressing unintentional threats, cognitive factors are given particular focus with Cybersecurity Awareness, Risk \& Protection Appraisal, and Security / Privacy Perception occuring often across assessments, alongside moderator indicators related to Awareness Raising, Formal Training, Age, and Gender. This suggests a strong tendency to conceptualize unintentional vulnerability primarily through insufficient awareness, limited security understanding, or inadequate decision-making processes during interaction with digital systems.

In contrast, studies addressing intentional threats demonstrate stronger emphasis on behavioral factors, in particular those associated with observable activity patterns, anomaly-related behavior, and behavioral monitoring approaches. Even though psychological factors, such as Personality Traits, appear in fewer studies in absolute terms compared to the unintentional category, their relative prevalence remains notable. However, it is worth mentioning that across both threat types the consideration of Personality Traits is identified in around 1/3 of the reviewed studies.

These observations suggest that intentional-threat-oriented approaches more frequently attempt to associate maliciousness tendency with deviant behavior and secondly with insider-related characteristics. At the same time, these studies showcase a focus on the Occupation indicator while incorporating comparatively fewer broader contextual or socio-cultural variables. This indicates that intentional maliciousness is often operationalized through behavioral deviation and user profiling rather than through deeper contextual, organizational, or environmental conditions that may dynamically influence intentional malicious cyber behavior.

Studies addressing both intentional and unintentional threats simultaneously, although limited in number, demonstrate broader factor coverage in relative terms than studies focusing on a single threat category. In addtion, it is observed that psychological factors, such as Emotion and Feelings and Self Perception, seem to have been considered more often in studies trying to address vulnerabilities of both threat types compared to studies aiming to address only a single threat category. Furthrmore, Age surfaces as the most considered moderator indicator, a trend that is not present in studies focusing on a single threat category. However, because only four studies belong to this category, findings remain inconclusive.

Overall, the comparison across threat categories indicates that the operationalization of vulnerability varies depending on the type of cyber threat addressed. However, despite these differences, the literature consistently prioritizes observable behavior while broader contextual, environmental, and socio-cultural conditions as well as deeper psychological mechanisms remain underrepresented across all threat categories.

\subsubsection{Assessment or Measurement Approaches}

\begin{figure}[H]
    \centering
    \includegraphics[width=0.9\linewidth]{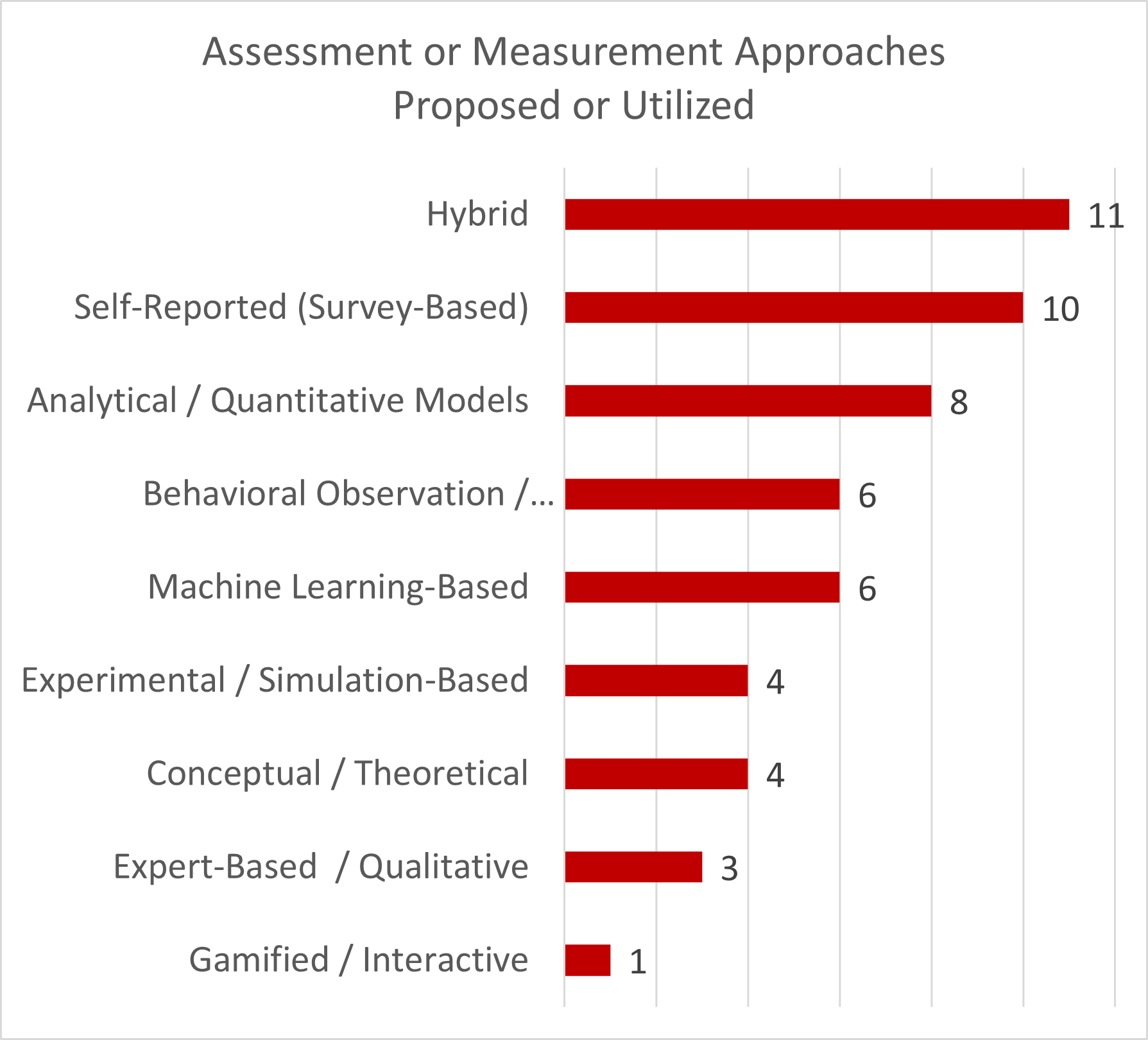}
    \caption{Assessment or Measurement Approaches Proposed or Utilized}
    \label{fig:fig19}
\end{figure}

 Regarding assessment or measurement approaches adopted across the reviewed studies, Figure \ref{fig:fig19} illustrates the distribution of proposed or utilized techniques. As depicted, the most common approach is hybrid methods, included in 11 studies, which indicates the interest to combine multiple techniques for the assessment or measurement of different human vulnerability factors in cybersecurity. This is followed by the self-reported (survey-based) approaches, with 10 studies, which remain frequently used, possibly due to their easy deployment and ability to gather user perceptions and behaviours. Analytical / quantitative models are also important, employed in 7 studies, which indicate efforts to standardize vulnerability assessment though measurable metrics and move beyond purely descriptive or perception-based assessment.

Following this, behavioral observation / monitoring and machine learning-based approaches, within 6 studies, highlight a focus in data-driven and automated analysis of user behavior. Some approaches include experiment / simulation-based methods and conceptual / theoretical approaches, with 4 studies each, while only one of the reviewed studies considers a gamified / interactive approach, indicating this method may be still emerging within the scope of HVA in cybersecurity.

\begin{figure}[H]
    \centering
    \includegraphics[width=0.9\linewidth]{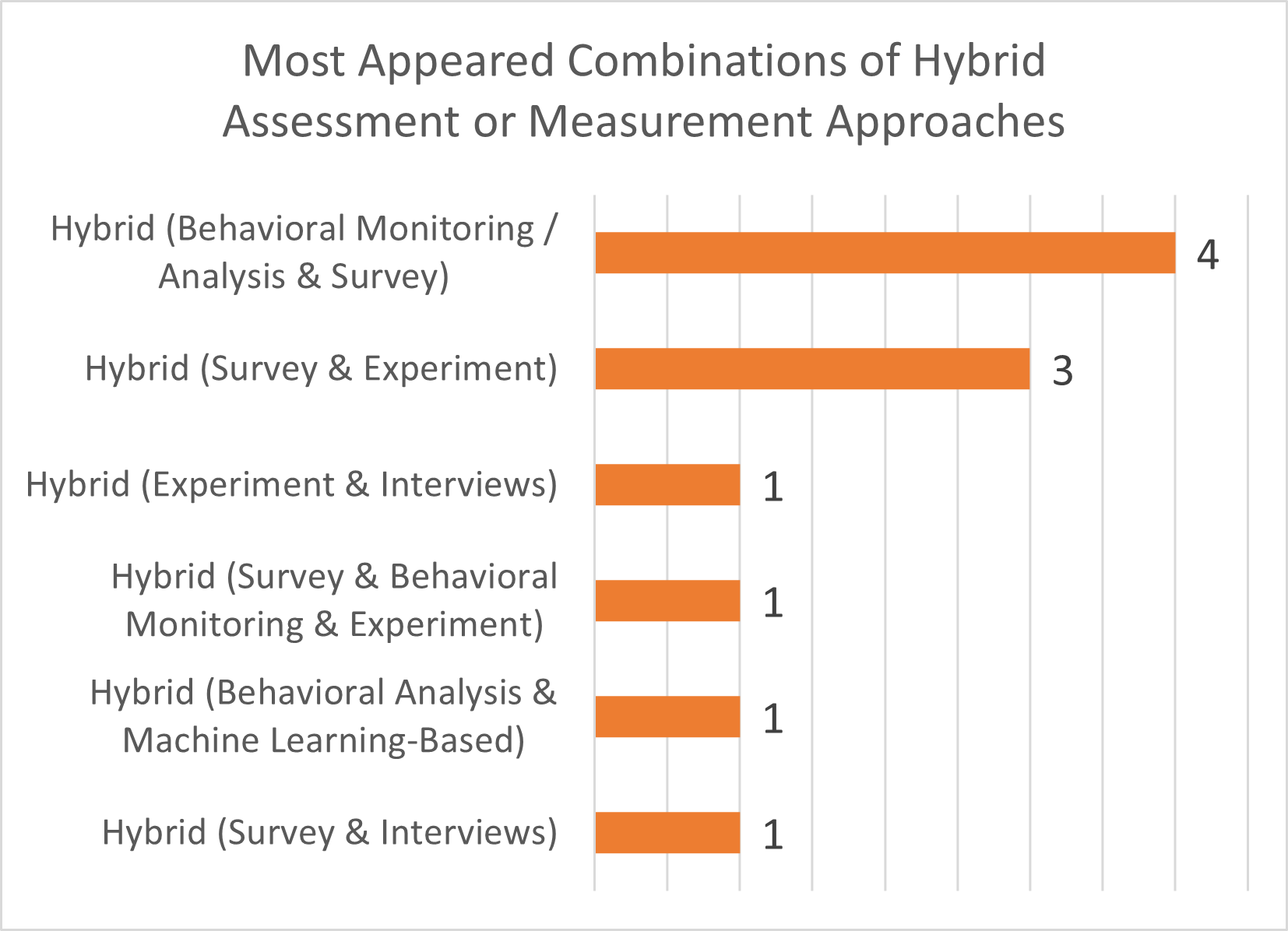}
    \caption{Most Appeared Combinations of Hybrid Assessment or Measurement Approaches}
    \label{fig:fig20}
\end{figure}

Considering the high number of hybrid approaches adopted across studies, Figure \ref{fig:fig20} provides information about the most appeared combinations. As illustrated, the most common combination is behavioral monitoring / analysis with survey-based methods, which have been identified in 4 studies and represent a preference to combine objective behavioral data with self-reported user insights. The second most common combination is survey-based approaches with experimental methods, which is described in 2 studies and indicates the need to complement subjective responses with controlled evaluation settings. The rest of the combinations appear only once, which suggests that there are diverse combinations but not frequently reused or yet fully explored across the literature.

\begin{figure}[H]
    \centering
    \includegraphics[width=0.9\linewidth]{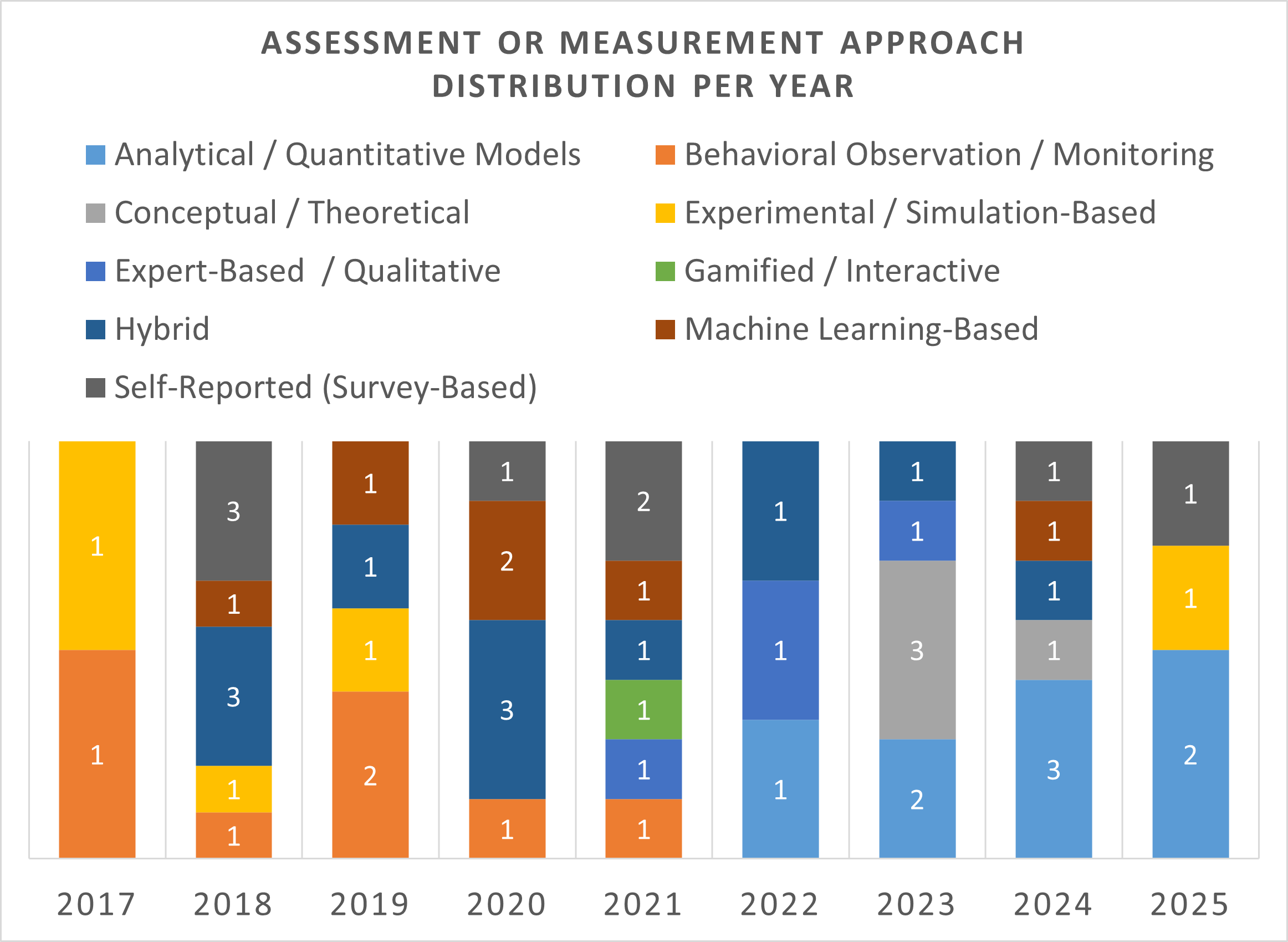}
    \caption{Assessment or Measurement Approach Distribution Per Year}
    \label{fig:fig21}
\end{figure}

Figure \ref{fig:fig21} compares the distribution of assessment and measurement approaches over time. It should be noted that the search was applied to the period 2017-2025, therefore the studies identified in 2014 and 2015, both of which are survey-based, were excluded from this diagram as they do not represent the entire research activity of those years. As depicted, early research (2017-2019) shows a consistent use of behavioral observation / monitoring, experimental / simulation-based, and self-reported (survey-based) approaches. Moreover, 2018 is a diverse year where many approaches have been considered, including the first use of hybrid and machine learning-based methods.

From 2020 and onwards, hybrid approaches become more common, indicating the need to combine multiple methodologies. Additionally, machine learning-based approaches have increased, which reflects the need for adoption of data-driven methods and a gradual shift toward behavior-centric and continuously observable assessment techniques. Self-reported (survey-based) approaches have decreased in recent years, but they are nevertheless included, which demonstrates their continued consideration, possibly due to their ease of use.

In recent years (2023-2025), research is centred on analytical / quantitative models, while conceptual / theoretical approaches are also included. This suggests the need for formalization and theoretical support. On the other hand, experimental / simulation-based and behavioral observation / monitoring approaches are less frequent, indicating a possible decrease in their use as standalone approaches. Furthermore, other approaches, such as the expert-based / qualitative and gamified / interactive methods, have not been widely used or mentioned throughout the years, which suggests that they are still understudied.

Overall, the figure showcases a shift from traditional, single-method approaches, such as surveys and experiments, to combined and data-driven approaches, such as hybrid, machine learning, and analytical methods. This reflects the need towards more complex and extensive measurement approaches in human vulnerability research in cybersecurity while highlighting that the field is still evolving conceptually and methodologically.

\subsubsection{Human Vulnerability Propagation \& Modelling}

\begin{figure}[H]
    \centering
    \includegraphics[width=0.9\linewidth]{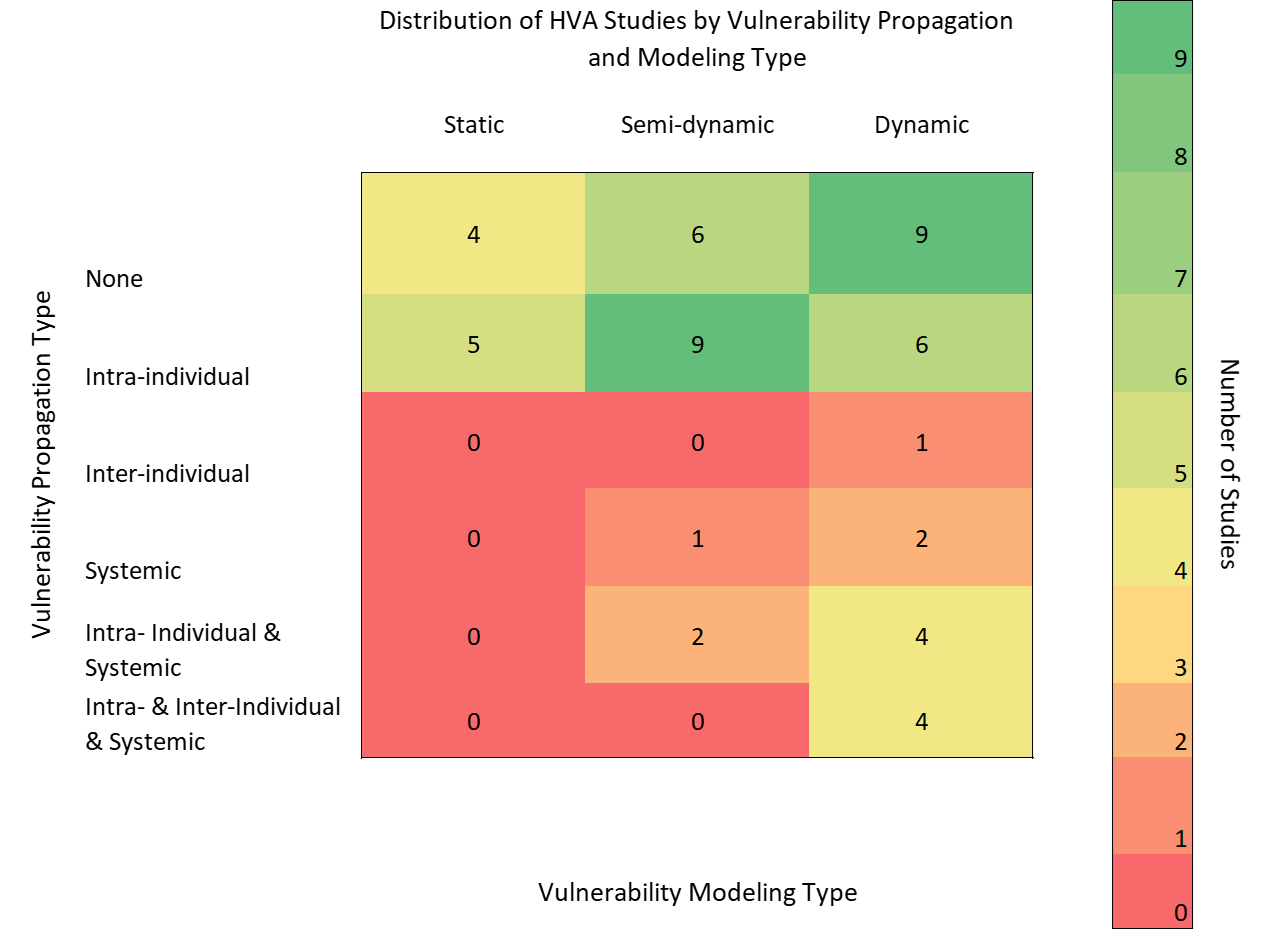}
    \caption{Distribution of HVA Studies by Vulnerability Propagation and Modeling Type}
    \label{fig:heatmap}
\end{figure}

Several patterns were revealed across the reviewed studies with regard to vulnerability propagation and modelling as well as regarding the relationship between them, which is depicted in Figure \ref{fig:heatmap}. Among the studies that operationalize propagation mechanisms, the most common appoach appears to be intra-individual propagation, identified in more than half of the studies, indicating that the literature recognizes that human vulnerability factors should not be assessed in isolation from one another, as the materialization of one may indicate the presense, increase, or decrease of another. However, inter-individual propagation remains exceptionally rare, while systemic propagation appears only in a limited subset of studies and is concentrated primarily within insider-threat, socio-technical, enterprise, or critical-infrastructure-oriented research.

Despite that, many studies still fall within the “None” propagation category, indicating that a sunstantial number of proposed HVA solutions, almost 1/3, still focus on an individual assessment level without investigating how vulnerabilities interact among individuals or propagate across socio-technical structures. In terms of modelling, it is also woth noting that even though dynamic approaches seem to be substantially more common than expected, they are still seldomly combined with vulnerability propagation. This highlights a disproportionate evolution in the field, as on one hand vulnerability assessment strongly moves towards approaches that acknowledge that it may evolve over time, but on the other hand it remains fragmented with regard to its social or relational propagation indicating that such effect mechanisms remain underrepresented.

A deeper analysis reveals clear methodological distinctions across modelling categories: static approaches are strongly associated with psychometric instruments and self-reported survey-based methods, whereas dynamic approaches correlate primarily with behavioral monitoring, simulation environments, UEBA systems, anomaly detection, and machine-learning-based techniques.

At the same time, multidimensionality does not necessarily imply propagation modelling, as several studies exhibit broad factor coverage while still lacking propagation mechanisms. In contrast, studies implementing multi-level propagation mechanisms tend to demonstrate comparatively richer conceptual structures overall, including broader factor coverage, larger moderator integration, and stronger socio-technical framing.

\subsection{Discussion: Trends, Gaps and Limitations}
The results of this SLR reflect a broader conceptual and methodological evolution in proposed HVA solutions in nearly the past decade. Early studies mostly addressed vulnerabilities related to unintentional threats considering cybersecurity awareness and psychometric constructs, focusing on static profiling and behavioral patterns through self-reported approaches that fall short in capturing actual behaviour under realistic operational conditions. Over the years, the literature started shifting towards identifying insider threats through UEBA systems, anomaly detection, and behavioral analysis based on machine learning approaches; however, without investigating the underlying vulnerability mechanisms that may indicate or affect the emergence of intentional threats. In parallel, behavioral monitoring has also been employed to oversee user actions continuously and at scale, but they provide limited visibility into internal cognitive, emotional, or psychological mechanisms. More recent research has attempted to address this gap through adaptive monitoring, context-aware systems, and continuous assessment approaches. This shift has also led to the emergence of hybrid vulnerability assessment or measurement methods combining behavioral observation with a narrow set of psychometric and contextual information. However, these approaches are still relatively rare, potentially due to higher complexity, privacy concerns, integration challenges, and dependence on continuous data collection.

Even though evolution in the field showcases a significant recognition of the multi-facet and continuously changing nature of human vulnerability, vulnerability itself is still largely understood through detection of observable user patterns, i.e., when risky behavior has already manifested, and remains far from a holistic consideration of the entire spectrum of human factors and moderator indicators. In addition, the more complex, underlying mechanisms, influencing risky actions, policy violations, browsing activity, or anomalous behavior patterns, remain underrepresented even in modern systems that adopt multi-factor assessment approaches. 

Moreover, in the ever-changing nature of AI-driven environments, there is little to no element of Human-AI interaction in the studies reviewed. Cognitive vulnerabilities related to automation bias, overtrust in AI systems, cognitive offloading, and AI-mediated manipulation are yet to be included in HVA approaches that need to be adapted to the evolving technological landscape. At the same time, the parallel consideration of both unintentional and intentional threats has been identified in a limited number of studies, while in most cases assessment is approached through different methodologies that do not intersect, indicating that accidental human errors and deliberate malicious actions continue to be treated as different phenomena instead of patters that could arise from common human, organisational, and contextual conditions. Behavioral dimensions and, in some cases, cognitive constructs have dominated the literature whereas psychological, contextual, and performance-related dimensions seem to lack similar attention in recent approaches, rendering proactive vulnerability identification an area still requiring investigation. 

Meanwhile, the conceptualization of vulnerability itself continues to be treated as a static trait without explicitly modelling long term susceptibility trajectories, changing psychological states, adaptive vulnerability states, peer influence, trust chain effects, collaborative susceptibility, and collective vulnerability. One-time questionnaires, fixed behavioral baselines, and pre-defined risk indicators that consider the assessment of individuals in isolation have constituted the most commonly adopted methods over the years. While influence between and among human factors within a single individual has been long recognized and more recent approaches tend to shift towards continuous and adaptive monitoring, there is still a lack of comprehensive considerations of vulnerability propagation across individuals and systems. Primarily, focus has been on tracking behavioral streams or anomaly patterns rather than modeling human vulnerability as an evolving and propagating socio-technical phenomenon, even though many cyber threats, such as social engineering and insider threats, inherently involve relational, organizational, and socially propagating dynamics. This observation is further supported by the fact that socio-cultural and environmental moderator indicators are largely underrepresented in current HVA approaches. Such variables may be indicative of how susceptibility manifests and vulnerability spreads among individuals and envrironments as well as of the ways intentional threats emerge across socio-technical systems.

In summary, the findings of this review indicate that human vulnerability in cybersecurity is still addressed in a fragmented way, with different studies focusing on isolated aspects of a very complex phenomenon. Future research thus would benefit from a shift away from isolated behavioral observations to continuously adaptive, psychologically grounded, context-aware, longitudinal and socially informed assessment ecosystems that are able to model vulnerability not only as an observable outcome, but also as an evolving human and socio-technical process.

\section{Conclusion}
\label{sec:Conclusion}

The aim of this study was to systematically review Human Vulnerability Assessment approaches in cybersecurity to investigate whether a single solution exists that considers the entire spectrum of human factors and additional variables affecting human vulnerability, related to the emergence of both unintentional and intentional threats while also considering vulnerability propagation mechanisms and continuous and dynamic vulnerability modelling. The paper synthesised the findings of 54 studies, from 2017 to 2025, across methods, models and assessment instruments and performed a quantitative and qualitive analysis to provide a comprehensive picture of the current landscape. In parallel, this study also proposed a structured Cybersecurity Human Factor Taxonomy classifying human vulnerability into the domains of Psychological, Cognitive, Behavioral, and Human Performance State, supported by moderating variable groups related to demographics, cybersecurity training, previous cyber incident experience, and socio-cultural and environmental context.

Common combinations, methodological trends, levels of holisticness, and differences in approaches were identified and delineated, indicating that the majority of current proposed HVA solutions is behavior-centric, where vulnerability is operationalized through users observable behavior, awareness factors, and anomaly detection. More complex psychological, contextual, relational, and dynamic aspects of vulnerability remain largely underrepresented. Although multidimensional and adaptive perspectives are increasingly adopted in the literature, most studies still consider vulnerability as a separate and relatively static individual attribute, rather than as a dynamic socio-technical phenomenon influenced by contextual factors, changes over time, and vulnerability propagation within and between individuals and systems. Impostantly, Human-AI interaction remains to be considered as a new and growing dimension of human vulnerability in the modern cybersecurity environment.

Despite the valuable insights gained, this review does not come without limitations. The proposed taxonomy is based on the literature reviewed, where the pool of human factors and moderator indicators has been synthesized based on the various studies that have investigated human vulnerability within cybersecurity contexts. It has been leveraged as a classification framework, supporting the analysis of the 54 identified studies of this SLR, and would benefit from further investigations about human vulnerability to deception and threat strategies in a broader manner, across other scientific domains such as criminology and social sciences. Following this investigation, additional sub-factors, manifestations, or emergent dimensions beyond those identified in the current study, within each vulnerability domain and moderator group, may be revealed and the resulting updated taxonomy would need to be formally validated or evaluated by experts in various domains.

Additionally, future research should focus on identifying relationships both through current literature and empirical research: i) among human factors to unravel internal effects that they have on each other, and ii) between moderator indicators and human factors to better understand how human vulnerability can be assessed and quantified in a realistic way. Vulnerability propagation should be further investigated as well, to identify how it can be conceptualized and operationalized at intra-individual, inter-individual, and systemic levels, especially in socio-technical settings where vulnerabilities may propagate or amplify each other over time.

Based on the insights gained through this SLR, the ultimate goal is to move towards the design and development of a holistic, dynamic, continuous, and propagation-aware HVA framework, incorporating psychometric models, behavioral monitoring, contextual reasoning, adaptive assessment mechanisms, and continuous monitoring tools. It is important to acknowledge also that this direction would require further investigation into the privacy, ethical, and governance issues of continuous human vulnerability assessment and monitoring to ensure that sensitive data are safeguarded and assessed individuals are not stigmatized.

\bibliographystyle{IEEEtran}
\bibliography{References.bib}

\begin{IEEEbiographynophoto}{Dimitra Papatsaroucha}
is a Cybersecurity Associate Researcher in the Department of Electrical and Computer Engineering at the Hellenic Mediterranean University and Senior Research Project Manager and Lab Team Leader at the Pasiphae R\&D Laboratory of HMU. She is currently pursuing a Ph.D. focused on Human Vulnerability Assessment and human-centered cybersecurity. Her research interests include cybersecurity, human factors in cybersecurity, privacy-enhancing technologies, secure data exchange, AI-enhanced threat detection, and secure digital infrastructures. She has contributed to more than 7 European research and innovation projects under Horizon 2020, Horizon Europe, and MSCA programmes, undertaking roles including Deputy Technical Coordinator and key scientific contributor in activities related to cybersecurity architectures, trusted digital ecosystems, privacy-preserving systems, and cyber resilience. She has co-authored more than 20 scientific publications and is actively involved in European proposal development, interdisciplinary research coordination, and stakeholder engagement within international R\&D initiatives. She has presented the findings of her work at international conferences and scientific events and she has contributed to the scientific community as a programme committee member and peer reviewer for international venues. Her work combines applied cybersecurity research, secure system design, and human-centered approaches for next-generation cybersecurity and resilient digital infrastructures.
\end{IEEEbiographynophoto}

\begin{IEEEbiographynophoto}{Stavroula Psaroudaki}
holds a Bachelor’s degree in Computer Science from the Democritus University of Thrace (DUTH) and is currently pursuing a Master of Science in Bioinformatics at the University of Crete. She is an Associate Researcher and Software Developer at the PASIPHAE Laboratory of the Hellenic Mediterranean University (HMU), actively contributing to European research and innovation projects under Horizon Europe and DIGITAL Europe. Her interests lie in the broader areas of computer science and include computational approaches to human vulnerability assessment, data analysis, methodological approaches for societal challenges, and bioinformatics.
\end{IEEEbiographynophoto}

\begin{IEEEbiographynophoto}{Eleftheria Vassilaki}
holds a Bsc in Business Administration from the Hellenic Mediterranean University and is currently pursuing her Msc in Tourism and Hospitality Management. She is a Research Associate in Pasiphae Lab, where she is a member of the Human Vulnerability Assessment research team, and participates in several European research projects. Her interests lie in expanding her knowledge of management practices, market research, and combining her backround in business administration with her presence in academia.
\end{IEEEbiographynophoto}

\begin{IEEEbiographynophoto}{Konstantina Pityanou}
received her B.Sc. degree in Informatics Engineering - Software Engineering from the Technological Educational Institute of Crete, in 2019. In 2022 she received her M.Sc. degree in Informatics Engineering from the Hellenic Mediterranean University (HMU) of Crete. She works as an Associate Researcher, Project Manager, and Software Developer at Pasiphae R\&D Laboratory of HMU, actively contributing to European research and innovation projects under Horizon Europe. Her interests include cybersecurity best practices, browser technologies, server-side technologies, database management, API development, and User Experience and User Interface design. She has co-authored more than 10 scientific publications and has presented the findings of her work at international conferences and scientific events. Her skills encompass both front-end and back-end development, reflecting her commitment to supporting innovative research efforts.
\end{IEEEbiographynophoto}

\begin{IEEEbiographynophoto}{Michail Alexandros Kourtis}
is a Researcher with the NCSR "Demokritos", working on NFV, SDN, 5G, and Cybersecurity. He has been involved in several EU research projects and the author of several publications in international journals and conferences (more than 90) in the respective fields. He has also collaborated as a contributor with ENISA on the "Threat Landscape for 5G Networks". He is the coordinator of four Horizon Europe projects in the domains of edge computing, cybersecurity certification, post-quantum cryptography, and governance.
\end{IEEEbiographynophoto}

\begin{IEEEbiographynophoto}{Ilias Politis}
is Research Associate at Industrial Systems Institute of ATHENA Research and Innovation Center, Greece. He received his BSc in Electronic Engineering from Queen Marry College London, UK in 2000, his MSc in Mobile and Personal Communications from King's College London, UK in 2001 and his PhD in Multimedia Communications from the University of Patras Greece in 2009. Since 2022 Dr Politis was a Senior Researcher with Secure Systems Labs of the University of Piraeus, Greece and has in the past worked as Senior Researcher at the Wireless Telecommunications Lab. of the Electrical and Computer Engineering at the University of Patras, Greece and the School of Science \& Technology in the Hellenic Open University, Greece. Dr. Politis has been actively involved in all phases of several H2020 and FP7 framework projects, as well as several national funded research projects. His research is focused on areas such as Future Internet and Next Generation and Time Sensitive Networking, ID management and Access control, Secure and Trust Networks, where he has published more than 90 peer reviewed journals and conference proceedings. He has been awarded a post-doctoral scholarship under the SIEMENS "Excellence" Program in the field of Telematic Applications by the State Scholarship Foundation (IKY), Greece for his PhD thesis. He is a member of the IEEE and the Technical Chamber of Greece.
\end{IEEEbiographynophoto}

\begin{IEEEbiographynophoto}{Evangelos K. Markakis}
is an Assistant Professor at the Department of Electrical and Computer Engineering of the Hellenic Mediterranean University and Principal Investigator of the Pasiphae R\&D Laboratory. His research focuses on cybersecurity, secure communications, post-quantum cryptography, quantum-resilient systems, secure software engineering, cyber-physical systems, and critical infrastructure protection. He has authored over 150 scientific publications and has participated in more than 60 European research and innovation projects, including FP5, FP6, FP7, H2020, and Horizon Europe actions. In several of these projects, he has served as Coordinator, Technical Coordinator, or key scientific contributor, leading activities in cybersecurity architecture, secure communications, trusted infrastructures, cyber resilience, and quantum-safe security. He has also contributed significantly to European proposal development, having helped raise over €100 million in competitive Horizon funding through proposals he has written or co-developed. His work combines academic research, applied cybersecurity engineering, and European project leadership, with strong emphasis on practical, interoperable, and standards-aware security solutions for public safety, healthcare, telecom, defence, and critical infrastructure domains.
\end{IEEEbiographynophoto}

\vfill

\end{document}